%% file: Contribution1.tex
\documentclass{PoS}
\usepackage{graphicx}
\usepackage{colordvi}
\usepackage{multirow}

\title{The spectrum of closed loops of fundamental flux in D=2+1 $SU(N)$ gauge theories.}

\ShortTitle{The spectrum of closed strings of fundamental flux in D=2+1}

\author{\speaker{Andreas Athenodorou} \\Rudolf Peierls Centre for Theoretical Physics \\
        University of Oxford\\
        E-mail: \email{athinoa@thphysics.ox.ac.uk}}

\author{Barak Bringoltz \\ Rudolf Peierls Centre for Theoretical Physics \\
        University of Oxford\\
        and \\
        Physics Department\\
        University of Washington, Seattle, WA 98195-1560, USA\\
        E-mail: \email{barak@phys.washington.edu}}

\author{Michael Teper  \\ Rudolf Peierls Centre for Theoretical Physics \\
        University of Oxford\\
    E-mail: \email{m.teper1@physics.ox.ac.uk}}

\abstract{We study the closed-string spectrum of SU($N$) gauge theories in the fundamental
 representation in 2+1 dimensions. We calculate the energies of the lowest lying $\sim30$ states using a large variety of operators characterised by the quantum numbers of parity and longitudinal momentum. We find that our results for the ground state are very well approximated by the Nambu-Goto (NG) predictions even for short strings. For the excited states,
we observe significant deviations
from the NG predictions only for very short
strings and they decrease rapidly with increasing
 string length.
Finally, we see that Nambu-Goto provides a much better description of our results than the effective string theoretical predictions. 
  We discuss the continuum and large-$N$ limits.}

\FullConference{The XXV International Symposium on Lattice Field Theory\\
         July 30-4 August 2007\\
         Regensburg, Germany}

\begin{document}
\section{Introduction}
\vskip -0.25cm
It is an old idea that large-$N$ QCD might be exactly reformulated as a string theory. Although this relation has never been made precise, the results obtained in lattice gauge theory leave little doubt that there is a connection between these two theories. One of these results that leads to such a conclusion is the energy spectrum of the flux-tube.
A flux-tube in $SU(N)$ gauge theories with length $l$ much larger than its width is expected to be described by an effective low energy string theory. This can be verified by studying the energy spectrum of the flux-tube and comparing it to the effective string theory prediction.

In this work we focus on the spectrum of the closed fundamental
flux-tube in pure $SU(N)$ gauge theories in $D=2+1$ dimensions. We
calculate the energies of the lowest lying $\sim$30 states. This
allows to extend the comparison with theoretical predictions to
states with more quantum numbers and a nontrivial degeneracy
structure. More precisely, we want to check whether the closed
flux-tube can be described by the Nambu-Goto (NG) string
model \cite{NGpapers}, and, if so, how good a description it is.

The flux-tubes that we study have lengths that range from $\sim
0.65 \,{\rm fm}$ to $\sim 2.60 \,{\rm fm}$ and the lattices we use
have spacings $a\simeq 0.04 \ {\rm fm}$ and $a\simeq 0.08 \ {\rm
fm}$ (Despite working in pure gauge theories and in $D=2+1$, we
define ${\rm 1 fm}$ through the convention $\sigma \equiv (440{\rm
MeV})^2$). The gauge groups that we study have $N=3, 6$. The use
of $N>3$ will suppress any mixing amplitudes (eg. glueball-string
mixing) that cannot be described by a simple low-energy effective
string theory model.

The study of confining flux-tubes with lattice techniques has been an active field for the past three decades, and we refer the reader to some recent papers \cite{strings-lattice-review}. For more details on the calculation and relevant references see our longer write-up \cite{ABM}.

\section{Lattice construction}
\vskip -0.25cm
We define our gauge theory on a three-dimensional periodic Euclidean space-time lattice with $L \times L_{\perp} \times L_T$ sites. It is important to mention that in order to minimize the finite volume effects, we increase $L_{\perp}$ and $L_T$ as we decrease the length of the flux-tube. For the calculation of the physical observables we perform Monte-Carlo simulations using the standard Wilson plaquette action:
\begin{eqnarray}
S_{\rm W}=\beta \sum_P \left[ 1- \frac{1}{N} {\rm Re} {\rm Tr}{U_P} \right],
\label{eq:SW}
\end{eqnarray}
The bare coupling $\beta$ is related to the dimensionful coupling $g^2$ through $\lim_{a\to 0}\beta={2N}/{ag^2}$. In the large--$N$ limit, the 't Hooft coupling $\lambda=g^2N$
is kept fixed, and so we must scale $\beta=2N^2/\lambda \propto N^2$
in order to keep the lattice spacing fixed.
The simulation we use mixes standard heat-bath and over-relaxation steps in the
ratio 1 : 4. These are implemented by updating $SU(2)$ subgroups using the Cabibbo-Marinari algorithm.

We have calculated the string spectrum for the case of $SU(3)$ with $a \simeq 0.04, 0.08 \ {\rm fm}$ and $SU(6)$ with $a \simeq 0.08 \ {\rm fm}$. For the case of $SU(3)$ and $a \simeq 0.08 \ {\rm fm}$, the string lengths ranged between $\sim 0.65 {\rm fm}$ and $\sim 2.60 {\rm fm}$ and for the two other cases between  $\sim 0.95 {\rm fm}$ and $\sim 2.0 {\rm fm}$.
\section{General strategy}
\vskip -0.25cm
\label{strategy}
 We calculate the energies of flux-tubes that are closed
around a spatial torus using the correlators of suitably smeared
Polyakov loops that wind once around the corresponding spatial
tori and that have vanishing transverse momentum. This is a
standard technique with smearing/blocking designed to enhance the
projection of our operators onto the physical states. We classify
our operators using the quantum numbers of transverse parity in
$P=\pm$ and winding momentum $q = 0,\pm 1, \pm 2,\dots$ in units
of $2\pi/l$.

For each combination of these quantum numbers we construct the full
correlation matrix of operators and use it to obtain best estimates for the
string states using a variational method
applied to the transfer matrix $\hat{T}=e^{-aH}$.

Since we are mostly interested in the excited states, it makes
sense to introduce transverse deformations in the simple Polyakov
loop. 
Using this procedure we can increase the number of different operators to 80-200 and construct Polyakov loops described by the quantum numbers of $P$ and $q$.
We present the
different paths used in the calculation in Table~\ref{Operators}.
\begin{table}[htb]
\centering{
\begin{tabular}[c]{||c||c||c||c||c||} \hline \hline
\includegraphics[height=2.5cm]{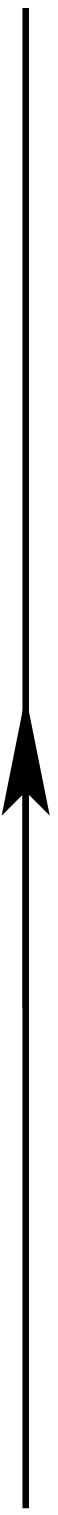}
&
\includegraphics[height=2.5cm]{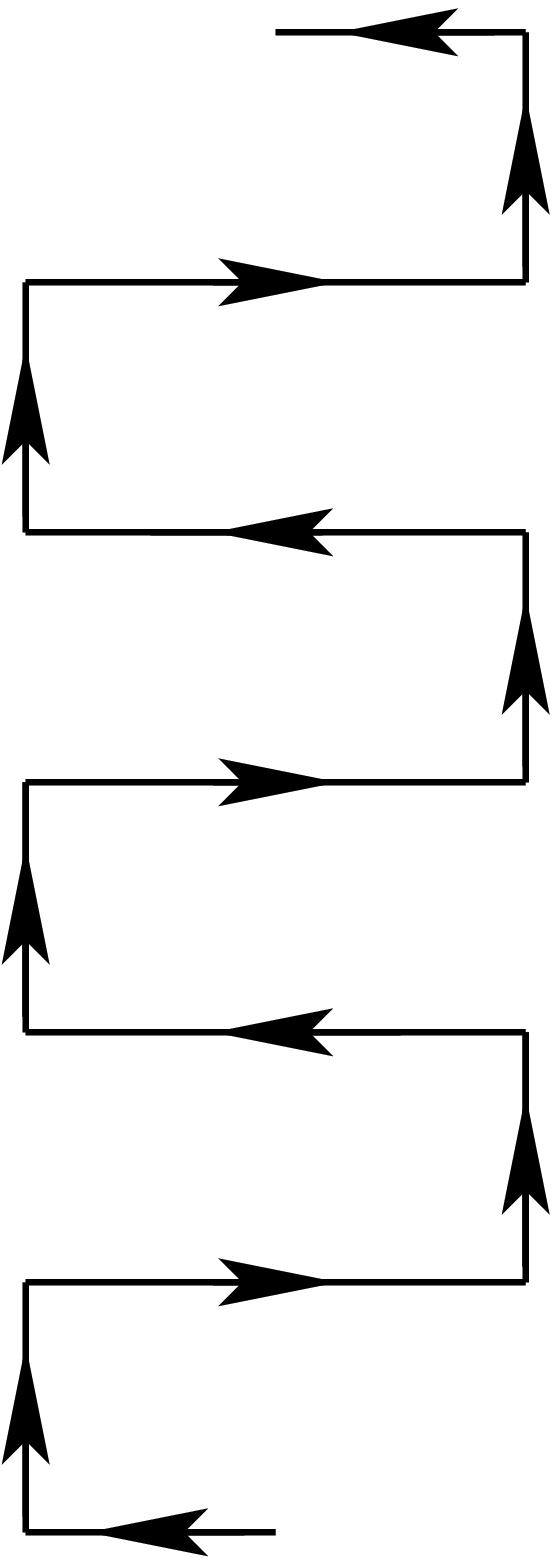}
&
\includegraphics[height=2.5cm]{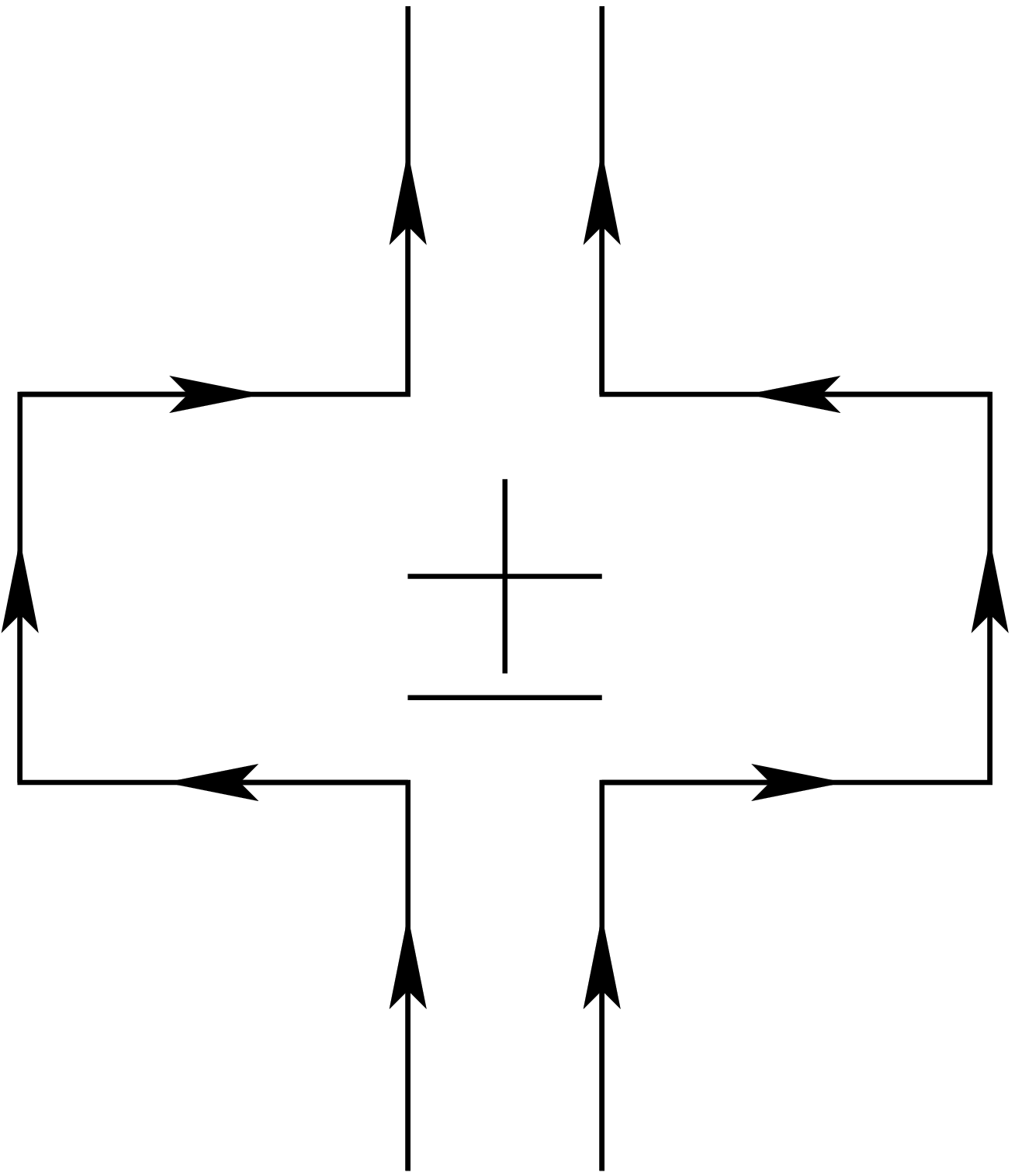}
&
\includegraphics[height=2.5cm]{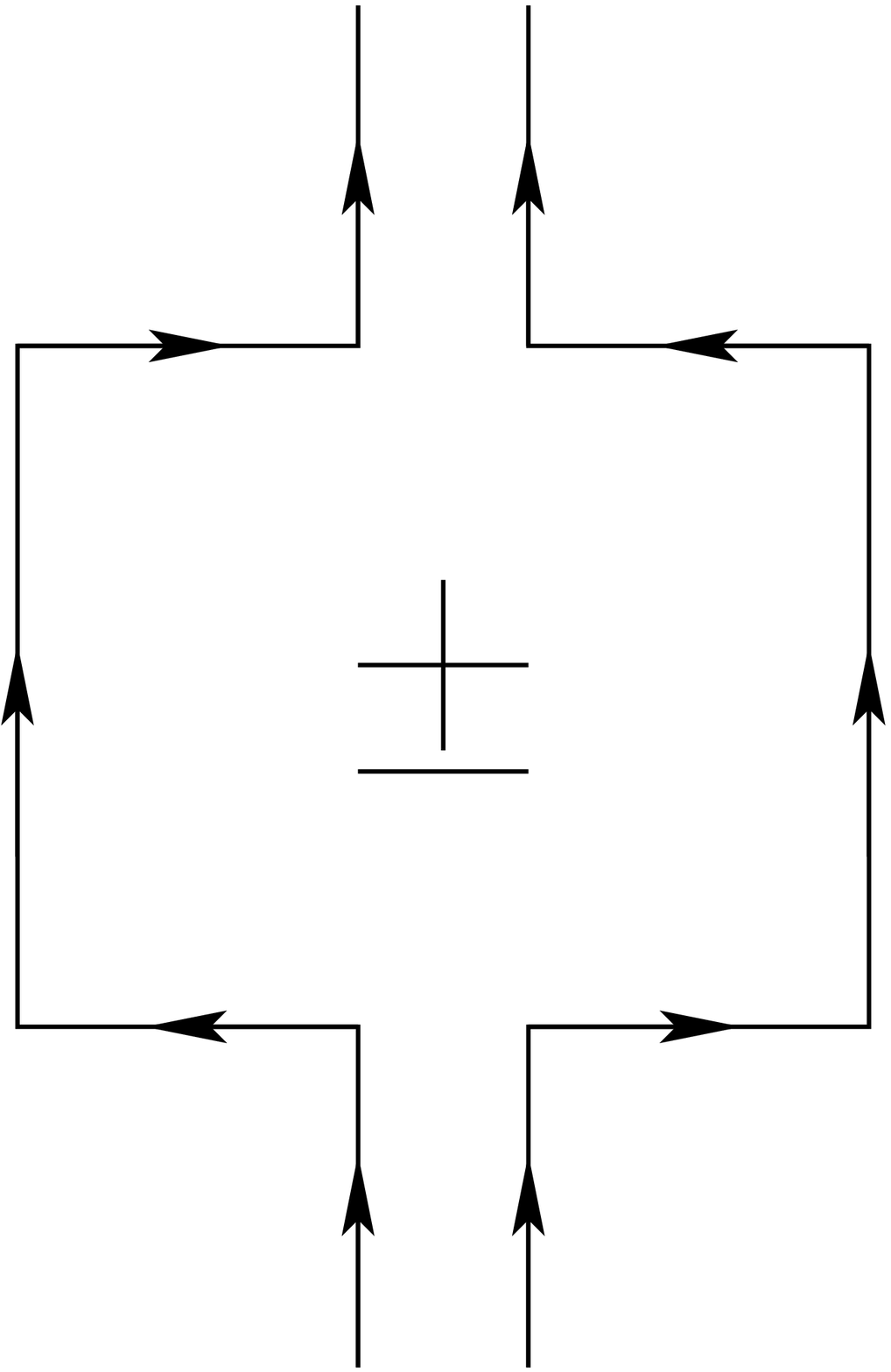}
&
\includegraphics[height=2.5cm]{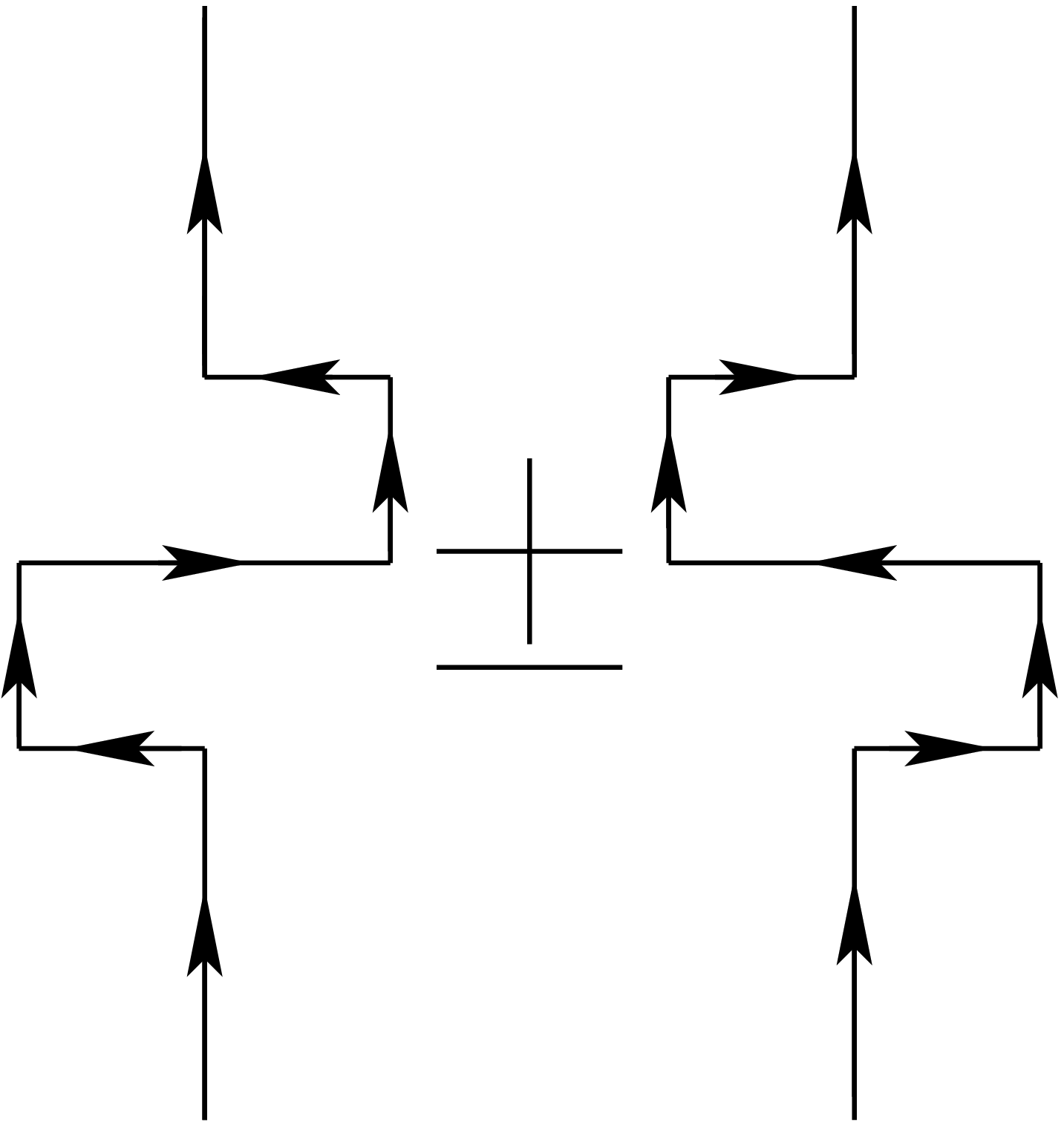} \\ \hline \hline

$1$ & $2$ & $3$ & $4$ &$5$ \\ \hline \hline

\includegraphics[height=2.5cm]{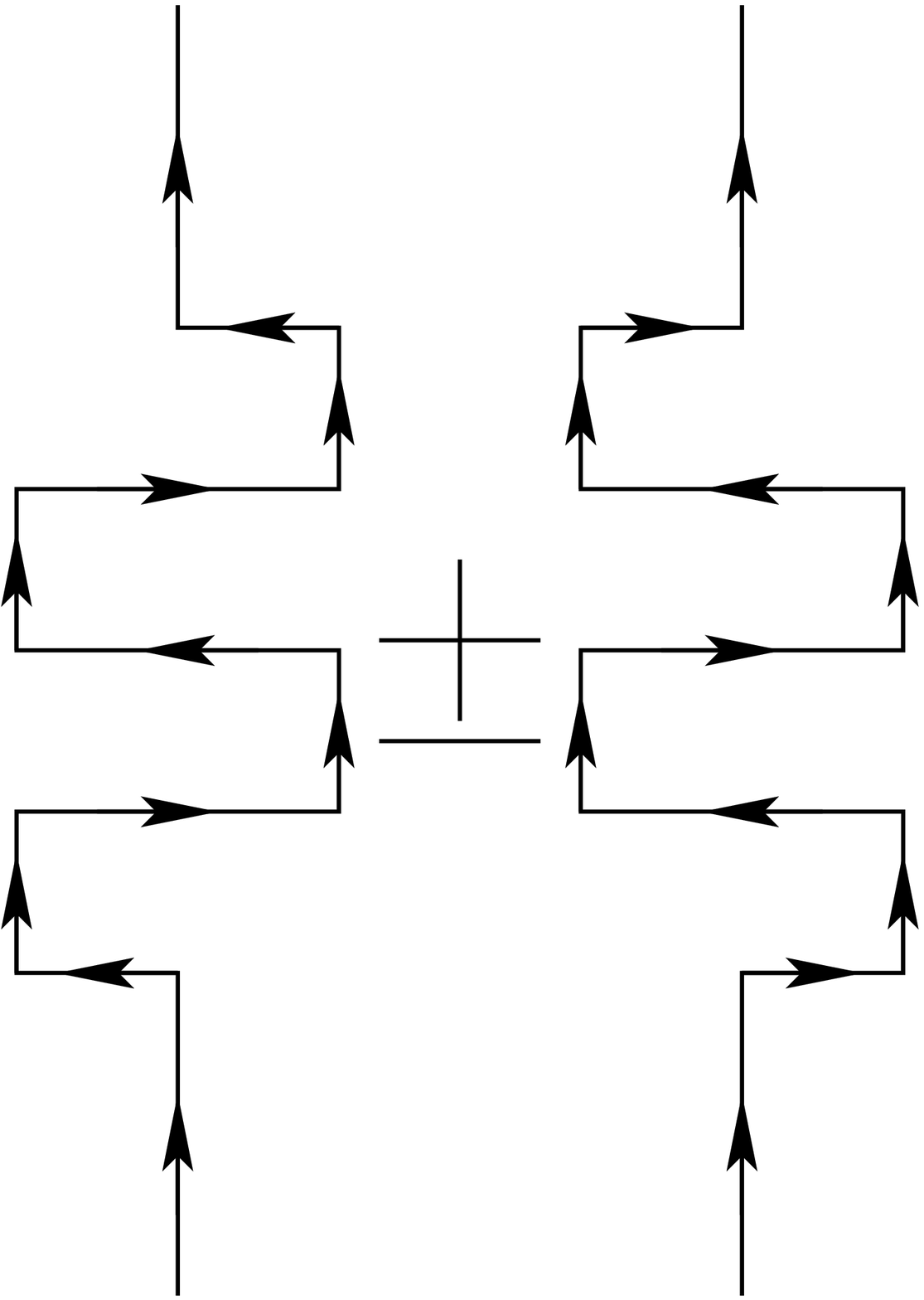}
&
\includegraphics[height=2.5cm]{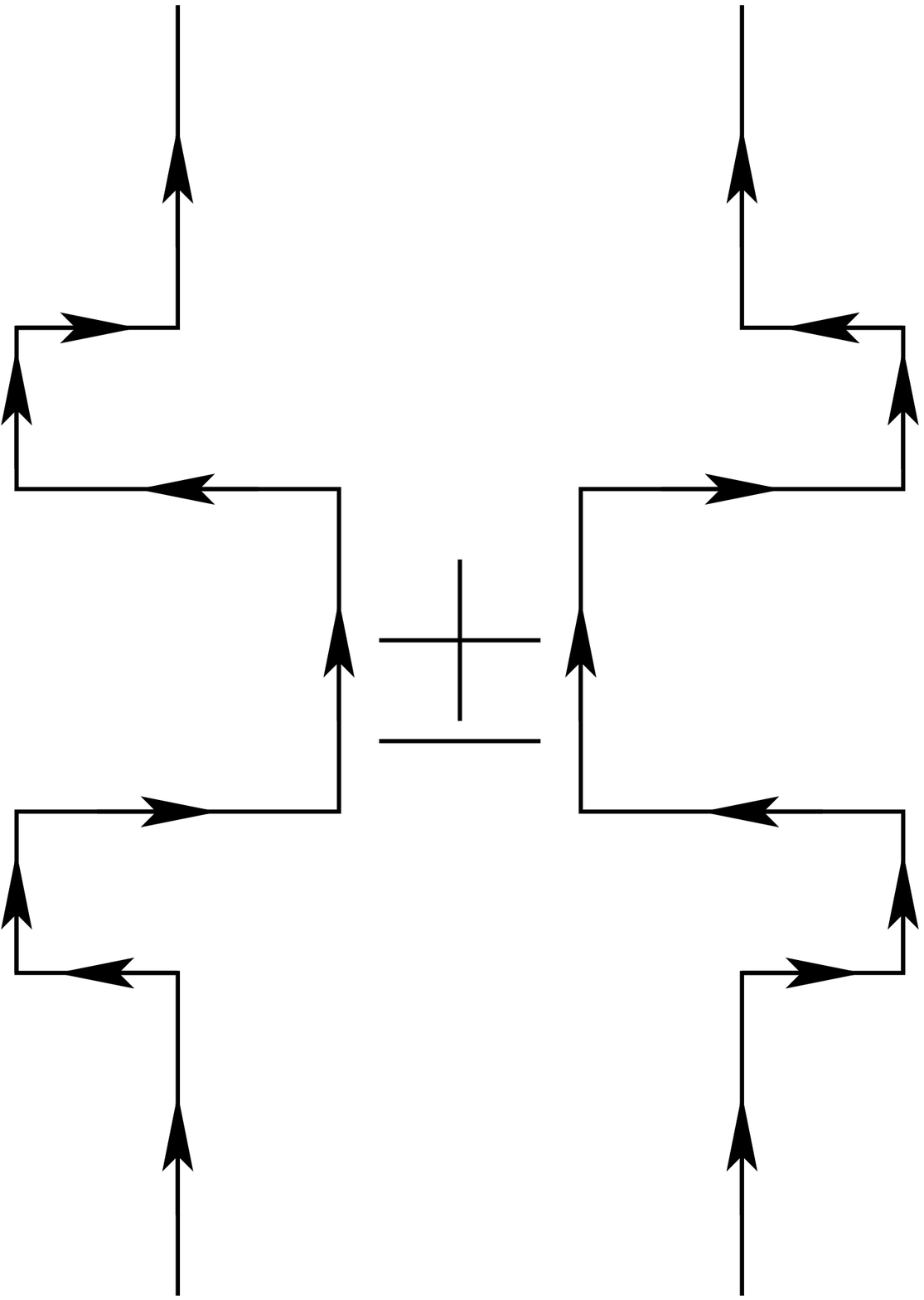}
&
\includegraphics[height=2.5cm]{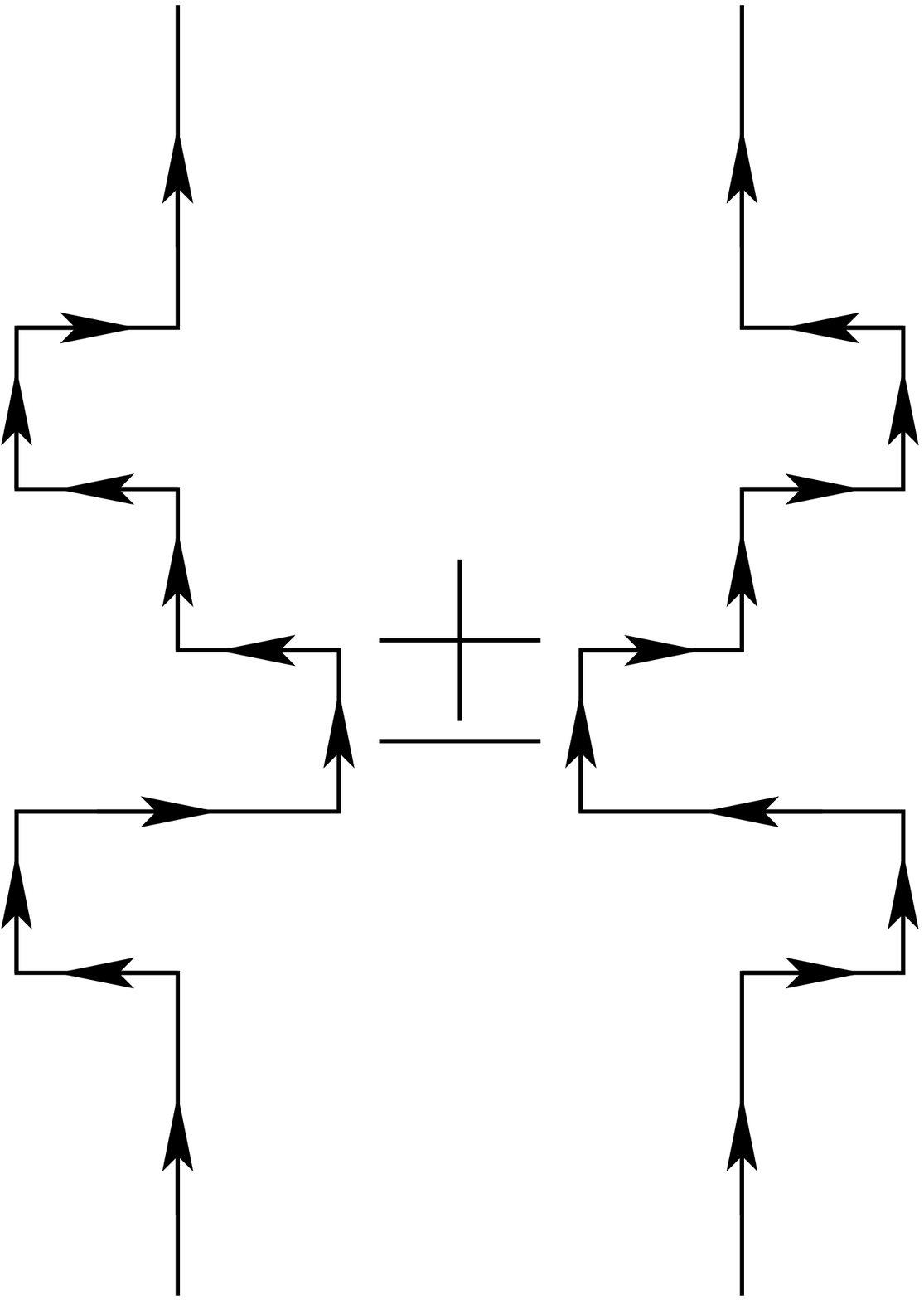}
&
\includegraphics[height=2.5cm]{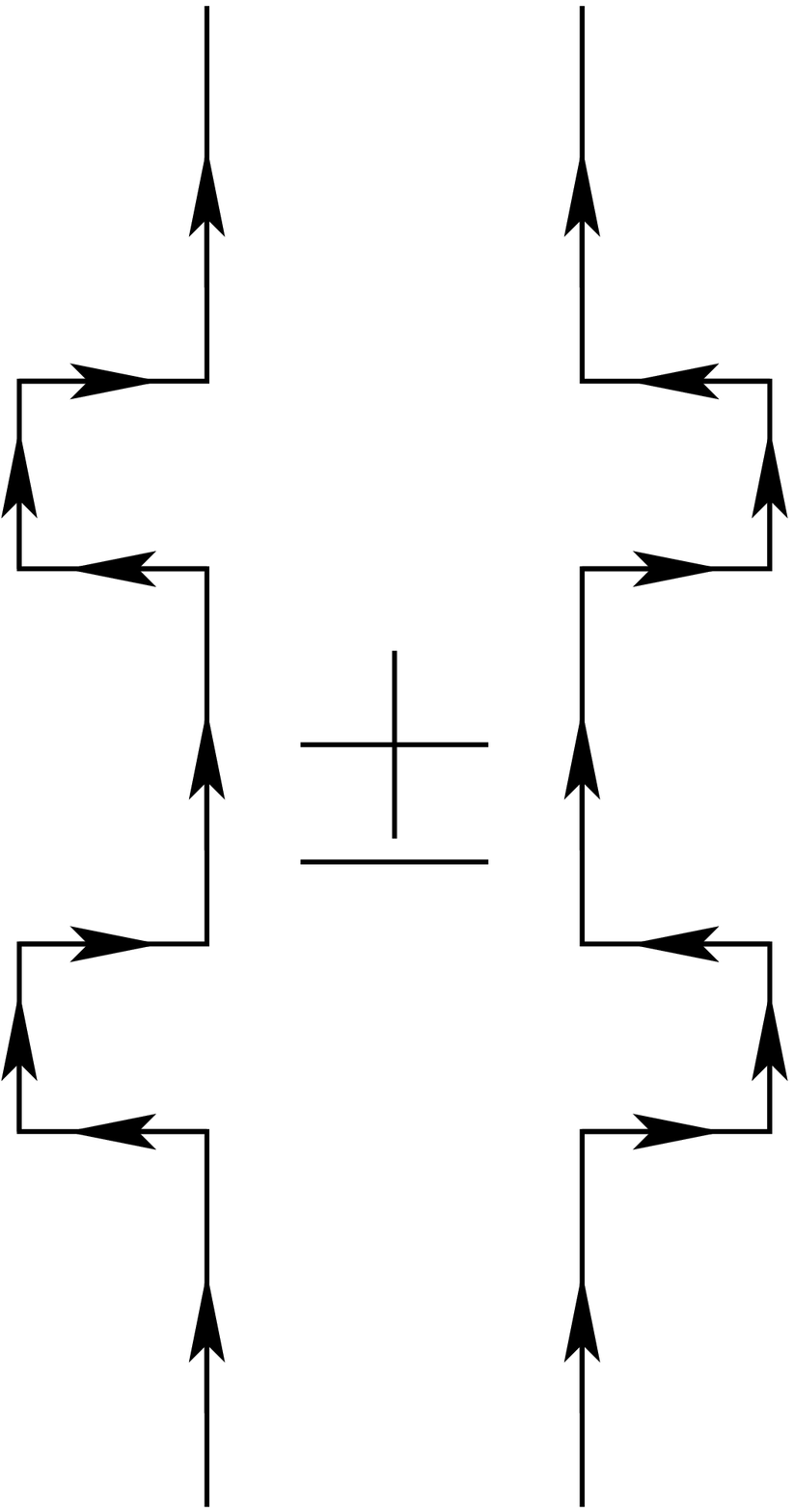}
&
\includegraphics[height=2.5cm]{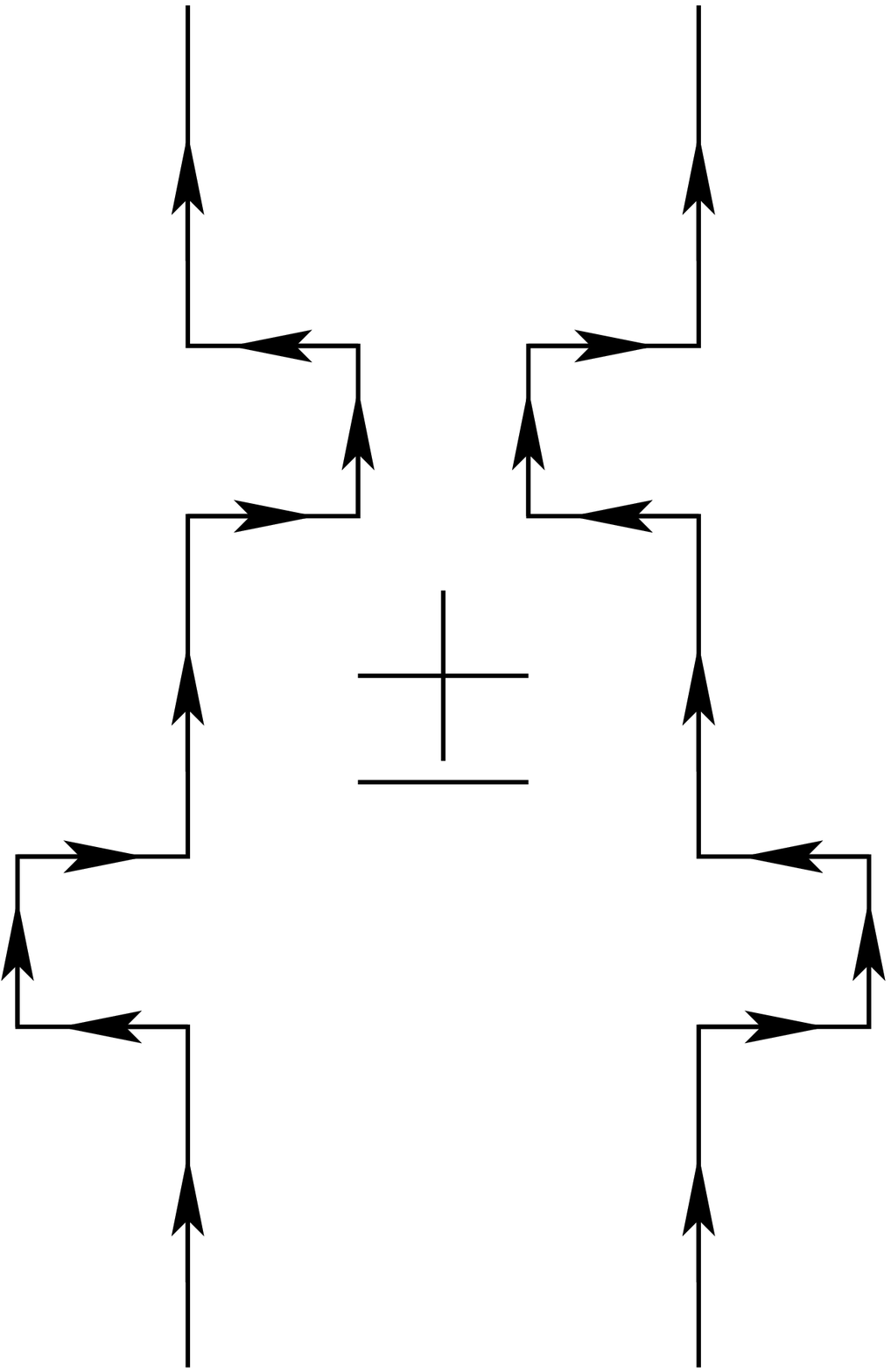} \\ \hline \hline

$6$ & $7$ & $8$ & $9$ & $10$ \\ \hline \hline

\includegraphics[height=1.75cm]{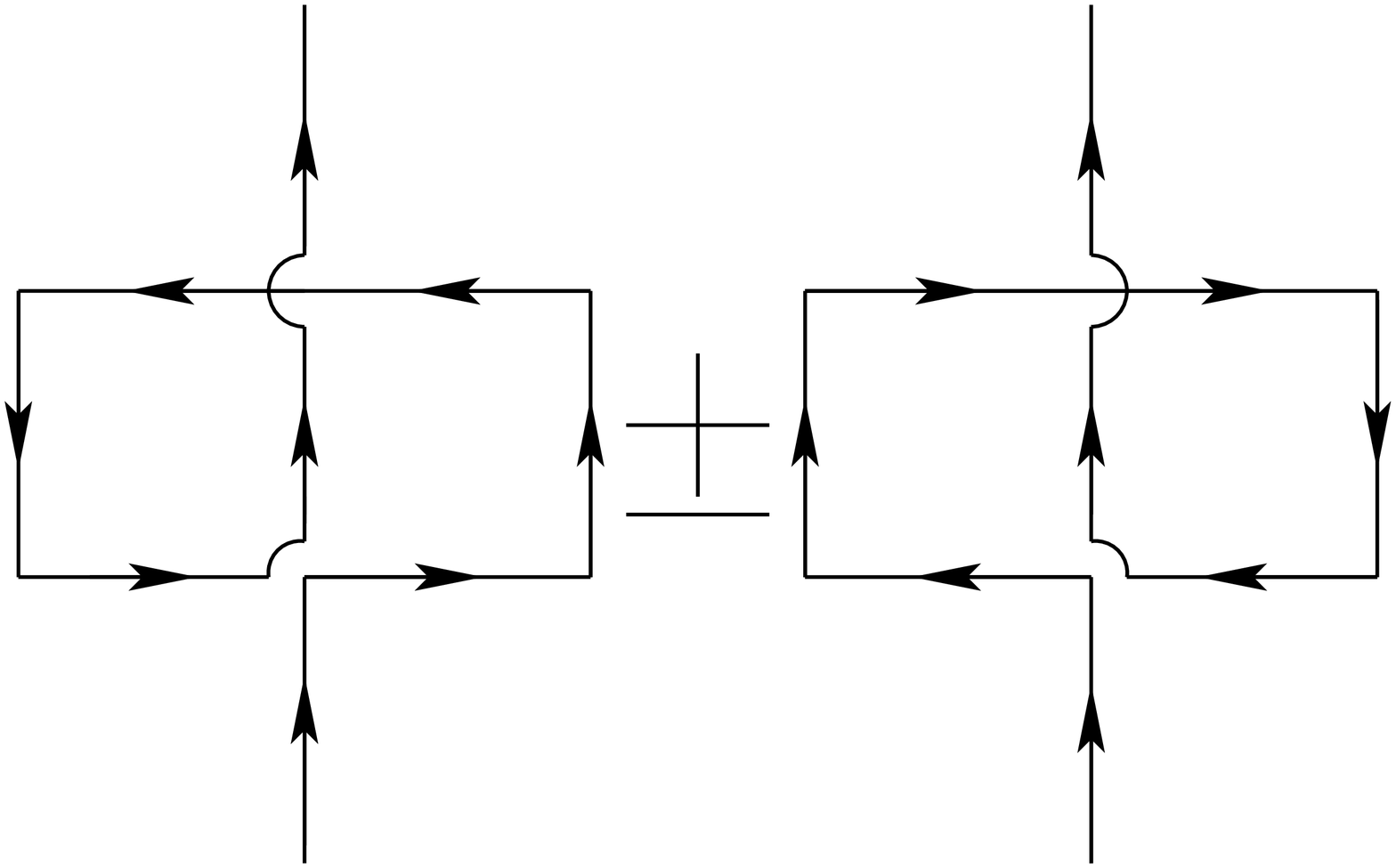}
&
\includegraphics[height=2cm]{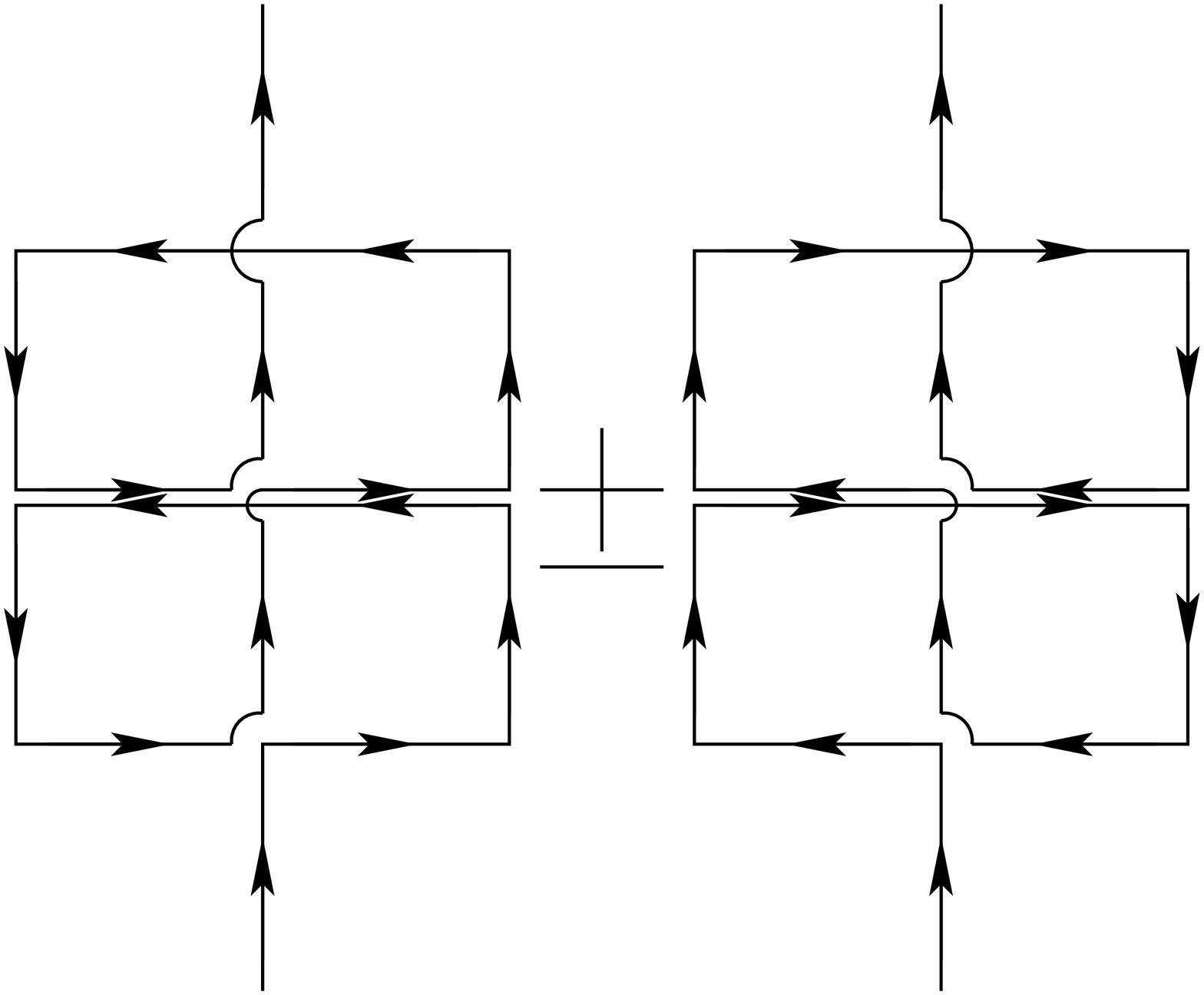}
&
\includegraphics[height=2cm]{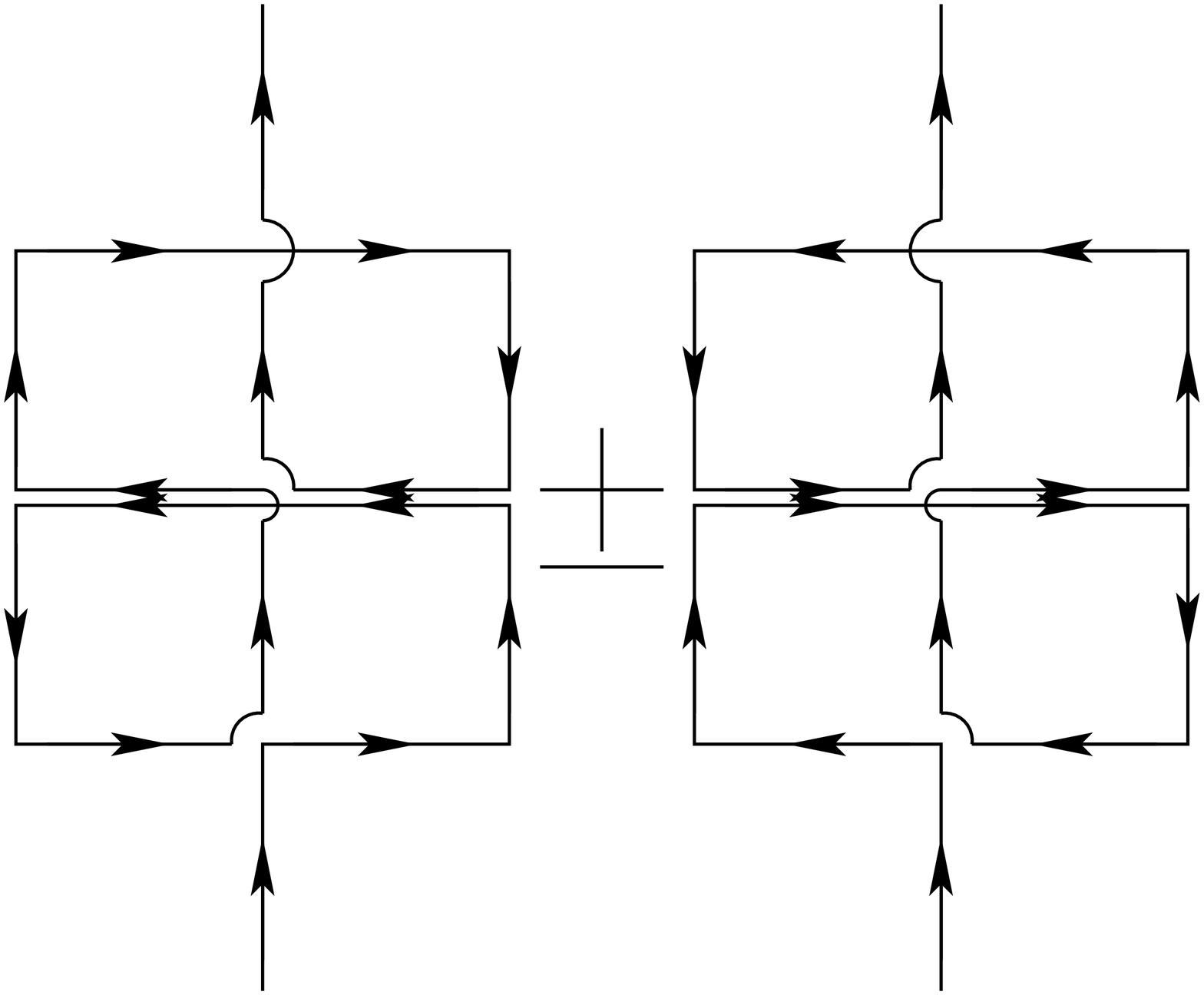}
&
\includegraphics[height=2cm]{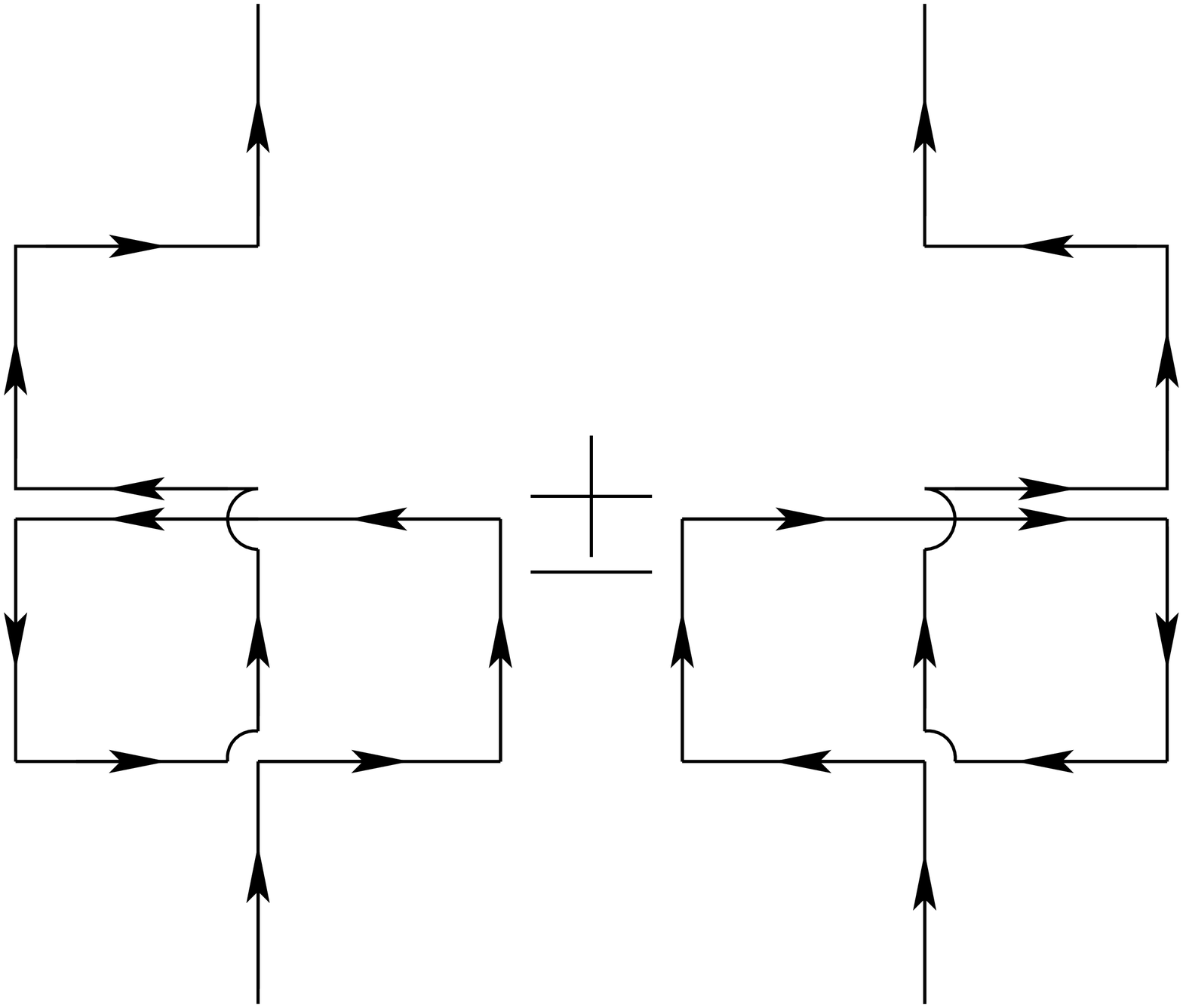}
&
\includegraphics[height=2cm]{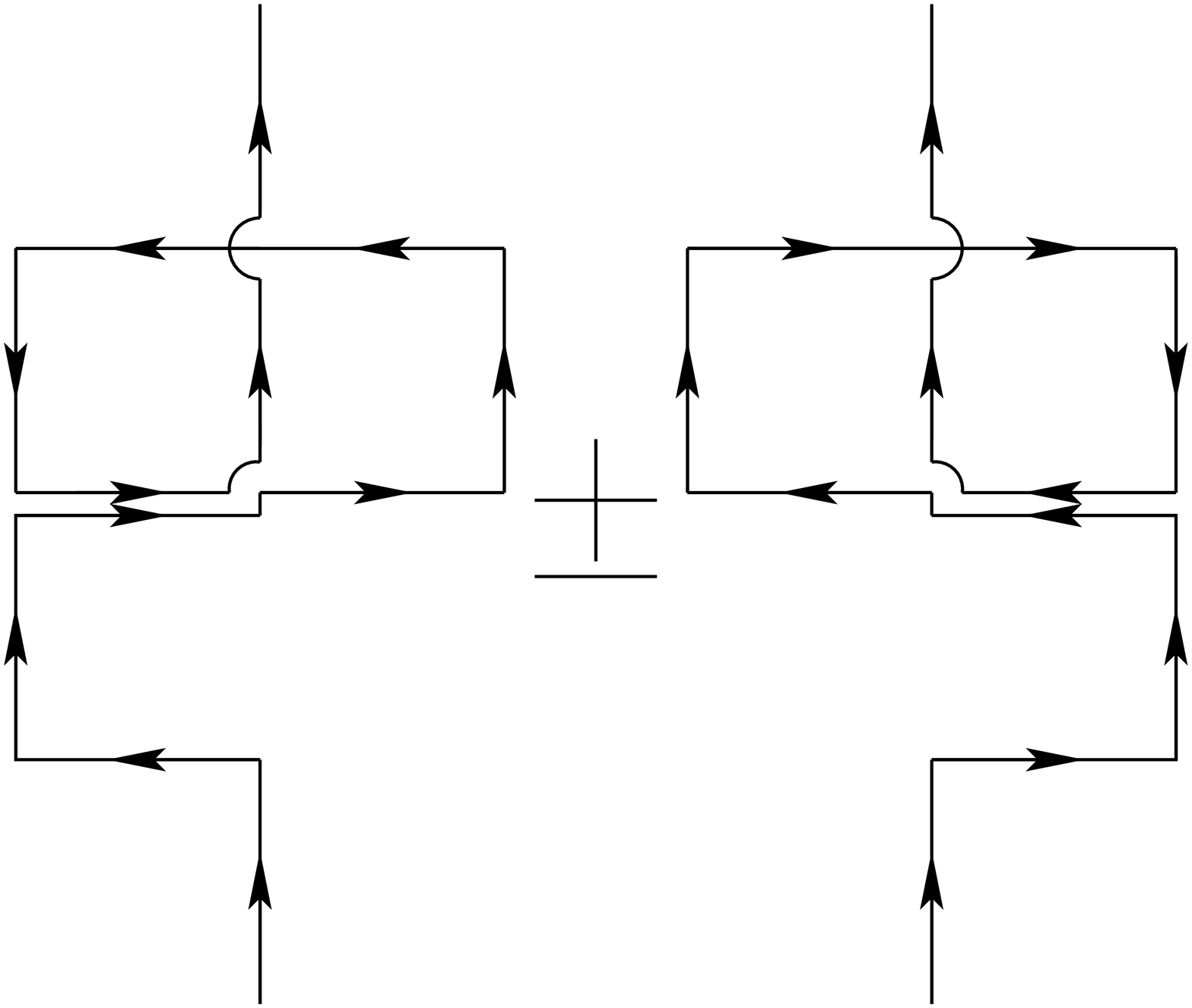} \\ \hline \hline

$11$& $12$ & $13$ & $14$ & $15$ \\ \hline \hline

\end{tabular}
}
\caption{The lattice paths used in the construction of Polyakov loops in this work. 
Our set of operators can be divided into three subsets: a. The
simple line operator (1) in several smearing/blocking
levels, b. The wave-like operator (2) whose number depends
upon $L$, $L_{\perp}$, and the smearing/blocking level, c. The
pulse-like operators (3-15) in several different
smearing/blocking levels.} \label{Operators}
\end{table}
\vskip -0.8cm
\section{Theoretical expectations: The spectrum of the Nambu-Goto String model}
\vskip -0.25cm
The action of the Nambu-Goto (NG) model \cite{NGpapers} is
the area of the
 worldsheet swept by the propagation of the string. This model is
quantum-mechanically inconsistent due to the Weyl anomaly ($D \ne
26$), but since this anomaly is suppressed for long strings
\cite{Olesen} it can still be considered as an effective low
energy model. The classical ground state of the model is the
 worldsheet configuration with the minimal area, and
 the transverse area fluctuation around this ground state constitute the quantum
 spectrum of the model.
 These fluctuations are usually referred to as left/right movers,
 depending on the momentum  that they carry in the direction parallel to the string axis.

The single string states can be characterised by the winding number $w$ which indicates how many times the string winds around the torus, by the occupation number $n_{L,(R)}(k)$ and the energy $k$ of the left and right movers and also
 by the center of mass momentum $\vec p_{\rm c.m.}$. By projecting to zero transverse momentum we are left only with
the momentum along the string axis which is quantized in units of
$2\pi q/l$ with $q=0,\pm 1,\pm 2,\dots$.
These quanta are not independent of $n_{L,R}$ and obey the level matching constraint
$N_L-N_R=qw$, where $N_{L(R)}$ enumerates the momentum contribution of the left(right) movers in a certain state as follows:
\begin{eqnarray}
N_L=\sum_{k>0} \sum_{n_L(k)>0} n_L(k)k \ \ \ \ {\rm and} \ \ \ \ N_R=\sum_{k'>0} \sum_{n_R(k')>0} n_R(k')k'
\end{eqnarray}
The string states can be characterised as irreducible representations of the $SO(D-2)$ group that rotates the spatial directions transverse to the string axis. In our $D=2+1$ case this group becomes the transverse parity with eigenvalues $P=(-1)^{\sum_{i=1} n_L(k_i) + \sum_{j=1} n_R(k'_j)}$.
Finally, the energy of a closed-string state described by the above quantum numbers for any $D$ is given by the following relation:
\begin{eqnarray}
E^2_{N_L,N_R,q,w} &=& \left(\sigma \,l w\right)^2 + 8 \pi \sigma
\left( \frac{N_L+N_R}2 - \frac{D-2}{24}\right) + \left(\frac{2\pi
q}{l}\right)^2.\label{NG0}
\end{eqnarray}
\section{Theoretical expectations: Effective string theories}
\vskip -0.25cm
Since in 2+1 dimensions the NG string is at best an effective low-energy string theory, it makes sense to generalise it  and write the most general form of an effective string action consistent with the symmetries of the flux-tube system. This was first done in the early eighties for the case of $w=1$ and $q=0$ in \cite{LW_old}, and the spectrum obtained for any $D$ was given by:
\begin{equation}
E_{n} = \sigma l + \frac{4\pi}{l} \left(n-\frac{D-2}{24}\right) + {\cal O}(1/l^2),\label{luscher0}
\end{equation}
where $n=0,1,2,...$. The second term in the above formula is known as the L\"uscher term and is expected to be universal. One can easily show that the NG model obeys this universality by expanding the square-root of Eq.~(\ref{NG0}) to leading order in $1/l$.

Recently, the work \cite{LW_old} was extended in \cite{LW_new}, where the authors used an open-closed string duality of the effective string theory. Using this duality they showed that for any $D$ the ${\cal O}(1/l^2)$ is absent from Eq.~(\ref{luscher0}), and that in $D=2+1$ the ${\cal O}(1/l^3)$ has a
 universal coefficient. Consequently, in $2+1$ dimensions
 Eq.~(\ref{luscher0}) is extended to:
\begin{equation}
E_{n} = \sigma l + \frac{4\pi}{l} \left(n-\frac{1}{24}\right) - \frac{8\pi^2}{\sigma \,l^3}
\left(n-\frac{1}{24}\right)^2 + {\cal O}(1/l^4),\label{luscher1}
\end{equation}
or in the equivalent form:
\begin{equation}
E^2_{n} = \left(\sigma l\right)^2 + 8\pi\sigma \left(n-\frac{1}{24}\right) + {\cal O}(1/l^3).\label{luscher2}
\end{equation}
The form Eq.~(\ref{luscher2}) is particularly convenient since the two first terms on the right hand side are the predictions of the NG model
that we find to be a very good approximation.

Motivated by this recent development, we decided to fit our data using the following
ansatz:
\begin{equation}
E^2_{\rm fit} = E^2_{NG} - \sigma
\frac{C_p}{\left(l\sqrt{\sigma}\right)^{p}}, \qquad p\ge
3\label{fit}
\end{equation}
where $E^2_{NG}$ is the Nambu Goto prediction given by Eq.~(\ref{NG0}) for $w=1, n=N_L=N_R$ and where $C_p$ are dimensionless coefficients that in general can depend on the quantum numbers of the state.
\section{Results}
\vskip -0.25cm
In this section we present our results from the calculation. The
results were obtained for $SU(3)$ with $\beta=21.00,40.00$  and
for $SU(6)$ with $\beta=90.00$. We have focused on strings with
$1.4 \stackrel{<}{_\sim} l\sqrt{\sigma}\stackrel{<}{_\sim} 5.5$.
All energies that we present were obtained from single cosh fits
to the correlation functions of our `best' operators - see
discussion in Section~\ref{strategy}.

We present the results in Figs.~\ref{Spectrum},\ref{mom}. The
lines are the predictions of the NG model. The string tensions
used for these predictions were extracted from the ground state
energies of the $q=0$ calculations with the use of the ansatz
Eq.~(\ref{fit}) for $p=3$. The fitting parameters are the
dimenionless $C_3$ and the lattice spacing in physical units
$a\sqrt{\sigma}$. We present these in
Table~\ref{fits-gs-partB}.
\begin{table}[htb]
\centering{
\begin{tabular}{|c|c|c|c|c|} \hline \hline
Gauge group  & $\beta$ & $a^2\sigma$ & $C_3$ & Confidence level \\ \hline \hline
\multirow{2}{0.45in}{$SU(3)$} & $\beta=21.00$  & $0.030258(26)$ & $0.160(21)$ & $69\%$    \\
 & $\beta=40.00$  & $0.007577(13)$ & $0.05(31)$ & $15\%$ \\ \hline
$SU(6)$ & $\beta=90.00$  & $0.029559(36)$ & $0.04(21)$   & $88\%$  \\ \hline
\end{tabular}
}
\caption{The parameters $a^2\sigma$ and $C_3$ in the fit.}
\label{fits-gs-partB}
\end{table}

According to our results the NG predictions are very good
approximations to the flux-tube spectrum and deviate from our data
only at the level of $\sim 2\%$ when
$l\sqrt{\sigma}\stackrel{>}{_\sim}4.2$. 
In contrast to that, a comparison of our data to Eq.~(\ref{luscher0}) and/or
Eq.(~\ref{luscher1}) fails even when
$l\sqrt{\sigma}\stackrel{>}{_\sim}4$. This is an important point
that tells us that the higher $O(1/l^4)$ terms in Eq.~(\ref{luscher1})
are significant for the excited states at these lengths, and that
they are captured quite well by Eq.~(\ref{NG0}) (to the level of $\sim
2\%$, a deviation that may or may not be due to a percent level
systematic errors that we did not aim to control for the excited
states). Finally, it is important to remark that the degeneracy
pattern predicted by the NG model is seen from our data. For
example, the second energy level is fourfold degenerate at
large-$l$. This degeneracy includes two positive parity states and
two negative parity states, and these start splitting
significantly once $l\sqrt{\sigma}\stackrel{<}{_\sim} 3$.
\begin{figure}
\centerline{ \scalebox{0.75}{\input{su3all2}  \put(-80,124){$n=0$}
\put(-80,168){$n=1$} \put(-80,204){$n=2$} \put(-193,170){LO}
\put(-200,130){NLO}} \ \scalebox{0.75}{\input{su6all2}
\put(-80,124){$n=0$} \put(-80,168){$n=1$} \put(-80,204){$n=2$}
\put(-193,170){LO} \put(-200,130){NLO}}} \caption{ The energies of
the first three energy levels divided by $\sqrt{\sigma}$ as
a function of $l \sqrt{\sigma}$. The three lines are the NG
predictions for the ground state and the excited states. The green
lines are the NG predictions expanded to leading order Eq. (5.1)
and next-to-leading order Eq. (5.2) for $n=1$. \underline{Left
panel:} We present the results for the case of $SU(3)$ for two
different values of $\beta$ (two different lattice spacings). For
$\beta=21$, positive parity states are presented in red and
negative parity states in blue. For $\beta=40$, positive parity
states are presented in cyan and negative parity states in pink.
\underline{Right panel:}  We present the results for the case of
$SU(6)$ with $\beta=90$. Positive parity states are presented in
red and negative parity states in blue.} \label{Spectrum}
\end{figure}
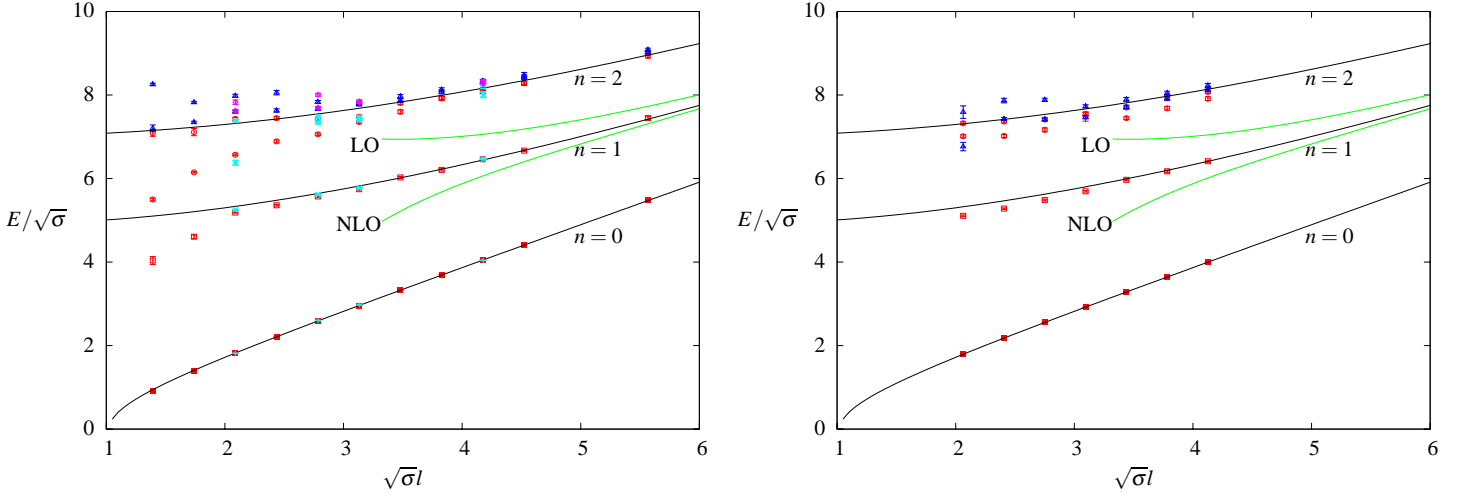
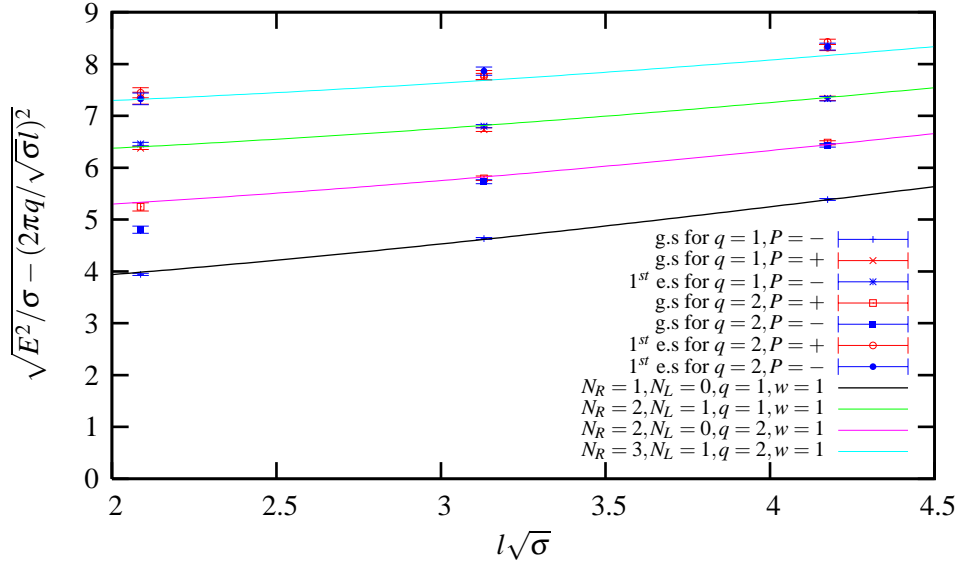
\begin{figure}[htb]
\centerline{
\scalebox{1.0}{\input{mom4}
}} \caption{$\sqrt{E^2/\sigma - \left( 2 \pi q/ \sqrt{\sigma} l
\right)^2}$ as a function of $l \sqrt{\sigma}$ for the ground
state and excited states of the non-zero
 momentum along the string direction. The
four lines present the NG predictions Eq. (4.2).
For $N_R=3, N_L=1$ and $q=2$, the expected degeneracy is three,
which agrees with our excited state data, i.e. two(one) states
with positive(negative) parity in red(blue).} \label{mom}
\end{figure}

Next, we performed fits of the energy of the excited states.
In the case of the first excited energy level, where there is only
one state per level, we used the fitting ansatz Eq.~(\ref{fit}).
In the case of the second excited energy level, where for each
parity there are two states, we fitted the difference between the
energies squared of these states. The fits showed that to
unambiguously determine the power $p$, and test the
L\"uscher-Weisz prediction of \cite{LW_new}, we need statistical
errors which are at least 2-3 times smaller than the ones our data
has, and a simultaneous control of any systematic errors that may
be important at the level of a few percents accuracy.

Finally, we move to the non-zero longitudinal momentum $q \ne 0$ calculation. In Section 4 we have mentioned that the number of left and right movers in the NG prediction is constrained by the level matching condition $N_L-N_R=qw$. The comparison of this prediction to our data is presented in Fig.~\ref{mom}, where we plot $\sqrt{E^2/\sigma - \left( 2 \pi q/ \sqrt{\sigma} l \right)^2}$ as a function of $l \sqrt{\sigma}$. It is clearly seen that our data is very well described by Eq.~(\ref{NG0}).
\section{Summary}
\vskip -0.25cm
We have calculated the energy spectrum of closed strings in the
fundamental representation of $SU(N)$ gauge theories in 2+1
dimensions. To perform this calculation we constructed a basis of
$\sim 80-200$ operators for each configuration of quantum numbers.
The use of this large basis of operators is
convenient for two reasons. Firstly, it is possible to
extract masses of excited states with high confidence since it increases the overlaps of our lattice operators onto the physical states (compared to what
we observe using the simple line operator in 5 smearing/blocking
levels). Secondly, it
enables us to study states with quantum numbers like the
transverse parity $P$ and the longitudinal momentum $q$.

Comparing our results to different theoretical predictions,
we find that the agreement with the NG prediction in
Eq.~(\ref{NG0}) is very good, including the expected degeneracy
pattern. This agreement is in striking contrast to what we find
when we simply compare to the L\"uscher term as in
Eq.~(\ref{luscher0}) or to the L\"uscher-Weisz prediction in
Eq.~(\ref{luscher1}). While for the ground states these describe
our data already at $l\sqrt{\sigma}\simeq 3$ where they are
indistinguishable from the full NG formula, for the excited states
the situation is completely different. In particular, whereas the
NG prediction works well for the first excited state already at
$l\sqrt{\sigma}\simeq 3$, the predictions of Eq.~(\ref{luscher0})
and Eq.~(\ref{luscher1}) are still very far from the data (see
Figure.~\ref{Spectrum}). This demonstrates that a confining
flux-tube can be described by a covariant string, with small or
moderate corrections down to very short lengths.
\section*{Acknowledgements}
The computations were carried out on PPARC and EPSRC funded computers in Oxford Theoretical Physics. AA acknowledges the support of the EC 6$^{th}$ Framework Programme Research and Training Network MRTN-CT-2004-503369. BB acknowledges the support of PPARC and thanks the hospitality of the Newton Institute for Mathematical Sciences at Cambridge, where some of this work was completed.

\end{document}

%% file: su3all2.tex
\begingroup%
\makeatletter%
\newcommand{\GNUPLOTspecial}{%
  \@sanitize\catcode`\%=14\relax\special}%
\setlength{\unitlength}{0.0500bp}%
\begin{picture}(7200,5040)(0,0)%
  {\GNUPLOTspecial{"
/gnudict 256 dict def
gnudict begin
%
%
/Color true def
/Blacktext false def
/Solid true def
/Dashlength 1 def
/Landscape false def
/Level1 false def
/Rounded false def
/TransparentPatterns false def
/gnulinewidth 5.000 def
/userlinewidth gnulinewidth def
/vshift -66 def
/dl1 {
  10.0 Dashlength mul mul
  Rounded { currentlinewidth 0.75 mul sub dup 0 le { pop 0.01 } if } if
} def
/dl2 {
  10.0 Dashlength mul mul
  Rounded { currentlinewidth 0.75 mul add } if
} def
/hpt_ 31.5 def
/vpt_ 31.5 def
/hpt hpt_ def
/vpt vpt_ def
Level1 {} {
/SDict 10 dict def
systemdict /pdfmark known not {
  userdict /pdfmark systemdict /cleartomark get put
} if
SDict begin [
  /Title (../../../../../../../LATTICE/POS/PLOTS/su3all2.tex)
  /Subject (gnuplot plot)
  /Creator (gnuplot 4.2 patchlevel 0)
  /Author (U-YOUR-4F60B1CEA3\\Andreas,S-1-5-21-2307956900-2123931569-774947135-1006)
  /CreationDate (Mon Sep 10 13:49:19 2007)
  /DOCINFO pdfmark
end
} ifelse
%
%
/M {moveto} bind def
/L {lineto} bind def
/R {rmoveto} bind def
/V {rlineto} bind def
/N {newpath moveto} bind def
/Z {closepath} bind def
/C {setrgbcolor} bind def
/f {rlineto fill} bind def
/vpt2 vpt 2 mul def
/hpt2 hpt 2 mul def
/Lshow {currentpoint stroke M 0 vshift R 
	Blacktext {gsave 0 setgray show grestore} {show} ifelse} def
/Rshow {currentpoint stroke M dup stringwidth pop neg vshift R
	Blacktext {gsave 0 setgray show grestore} {show} ifelse} def
/Cshow {currentpoint stroke M dup stringwidth pop -2 div vshift R 
	Blacktext {gsave 0 setgray show grestore} {show} ifelse} def
/UP {dup vpt_ mul /vpt exch def hpt_ mul /hpt exch def
  /hpt2 hpt 2 mul def /vpt2 vpt 2 mul def} def
/DL {Color {setrgbcolor Solid {pop []} if 0 setdash}
 {pop pop pop 0 setgray Solid {pop []} if 0 setdash} ifelse} def
/BL {stroke userlinewidth 2 mul setlinewidth
	Rounded {1 setlinejoin 1 setlinecap} if} def
/AL {stroke userlinewidth 2 div setlinewidth
	Rounded {1 setlinejoin 1 setlinecap} if} def
/UL {dup gnulinewidth mul /userlinewidth exch def
	dup 1 lt {pop 1} if 10 mul /udl exch def} def
/PL {stroke userlinewidth setlinewidth
	Rounded {1 setlinejoin 1 setlinecap} if} def
/LCw {1 1 1} def
/LCb {0 0 0} def
/LCa {0 0 0} def
/LC0 {1 0 0} def
/LC1 {0 1 0} def
/LC2 {0 0 1} def
/LC3 {1 0 1} def
/LC4 {0 1 1} def
/LC5 {1 1 0} def
/LC6 {0 0 0} def
/LC7 {1 0.3 0} def
/LC8 {0.5 0.5 0.5} def
/LTw {PL [] 1 setgray} def
/LTb {BL [] LCb DL} def
/LTa {AL [1 udl mul 2 udl mul] 0 setdash LCa setrgbcolor} def
/LT0 {PL [] LC0 DL} def
/LT1 {PL [4 dl1 2 dl2] LC1 DL} def
/LT2 {PL [2 dl1 3 dl2] LC2 DL} def
/LT3 {PL [1 dl1 1.5 dl2] LC3 DL} def
/LT4 {PL [6 dl1 2 dl2 1 dl1 2 dl2] LC4 DL} def
/LT5 {PL [3 dl1 3 dl2 1 dl1 3 dl2] LC5 DL} def
/LT6 {PL [2 dl1 2 dl2 2 dl1 6 dl2] LC6 DL} def
/LT7 {PL [1 dl1 2 dl2 6 dl1 2 dl2 1 dl1 2 dl2] LC7 DL} def
/LT8 {PL [2 dl1 2 dl2 2 dl1 2 dl2 2 dl1 2 dl2 2 dl1 4 dl2] LC8 DL} def
/Pnt {stroke [] 0 setdash gsave 1 setlinecap M 0 0 V stroke grestore} def
/Dia {stroke [] 0 setdash 2 copy vpt add M
  hpt neg vpt neg V hpt vpt neg V
  hpt vpt V hpt neg vpt V closepath stroke
  Pnt} def
/Pls {stroke [] 0 setdash vpt sub M 0 vpt2 V
  currentpoint stroke M
  hpt neg vpt neg R hpt2 0 V stroke
 } def
/Box {stroke [] 0 setdash 2 copy exch hpt sub exch vpt add M
  0 vpt2 neg V hpt2 0 V 0 vpt2 V
  hpt2 neg 0 V closepath stroke
  Pnt} def
/Crs {stroke [] 0 setdash exch hpt sub exch vpt add M
  hpt2 vpt2 neg V currentpoint stroke M
  hpt2 neg 0 R hpt2 vpt2 V stroke} def
/TriU {stroke [] 0 setdash 2 copy vpt 1.12 mul add M
  hpt neg vpt -1.62 mul V
  hpt 2 mul 0 V
  hpt neg vpt 1.62 mul V closepath stroke
  Pnt} def
/Star {2 copy Pls Crs} def
/BoxF {stroke [] 0 setdash exch hpt sub exch vpt add M
  0 vpt2 neg V hpt2 0 V 0 vpt2 V
  hpt2 neg 0 V closepath fill} def
/TriUF {stroke [] 0 setdash vpt 1.12 mul add M
  hpt neg vpt -1.62 mul V
  hpt 2 mul 0 V
  hpt neg vpt 1.62 mul V closepath fill} def
/TriD {stroke [] 0 setdash 2 copy vpt 1.12 mul sub M
  hpt neg vpt 1.62 mul V
  hpt 2 mul 0 V
  hpt neg vpt -1.62 mul V closepath stroke
  Pnt} def
/TriDF {stroke [] 0 setdash vpt 1.12 mul sub M
  hpt neg vpt 1.62 mul V
  hpt 2 mul 0 V
  hpt neg vpt -1.62 mul V closepath fill} def
/DiaF {stroke [] 0 setdash vpt add M
  hpt neg vpt neg V hpt vpt neg V
  hpt vpt V hpt neg vpt V closepath fill} def
/Pent {stroke [] 0 setdash 2 copy gsave
  translate 0 hpt M 4 {72 rotate 0 hpt L} repeat
  closepath stroke grestore Pnt} def
/PentF {stroke [] 0 setdash gsave
  translate 0 hpt M 4 {72 rotate 0 hpt L} repeat
  closepath fill grestore} def
/Circle {stroke [] 0 setdash 2 copy
  hpt 0 360 arc stroke Pnt} def
/CircleF {stroke [] 0 setdash hpt 0 360 arc fill} def
/C0 {BL [] 0 setdash 2 copy moveto vpt 90 450 arc} bind def
/C1 {BL [] 0 setdash 2 copy moveto
	2 copy vpt 0 90 arc closepath fill
	vpt 0 360 arc closepath} bind def
/C2 {BL [] 0 setdash 2 copy moveto
	2 copy vpt 90 180 arc closepath fill
	vpt 0 360 arc closepath} bind def
/C3 {BL [] 0 setdash 2 copy moveto
	2 copy vpt 0 180 arc closepath fill
	vpt 0 360 arc closepath} bind def
/C4 {BL [] 0 setdash 2 copy moveto
	2 copy vpt 180 270 arc closepath fill
	vpt 0 360 arc closepath} bind def
/C5 {BL [] 0 setdash 2 copy moveto
	2 copy vpt 0 90 arc
	2 copy moveto
	2 copy vpt 180 270 arc closepath fill
	vpt 0 360 arc} bind def
/C6 {BL [] 0 setdash 2 copy moveto
	2 copy vpt 90 270 arc closepath fill
	vpt 0 360 arc closepath} bind def
/C7 {BL [] 0 setdash 2 copy moveto
	2 copy vpt 0 270 arc closepath fill
	vpt 0 360 arc closepath} bind def
/C8 {BL [] 0 setdash 2 copy moveto
	2 copy vpt 270 360 arc closepath fill
	vpt 0 360 arc closepath} bind def
/C9 {BL [] 0 setdash 2 copy moveto
	2 copy vpt 270 450 arc closepath fill
	vpt 0 360 arc closepath} bind def
/C10 {BL [] 0 setdash 2 copy 2 copy moveto vpt 270 360 arc closepath fill
	2 copy moveto
	2 copy vpt 90 180 arc closepath fill
	vpt 0 360 arc closepath} bind def
/C11 {BL [] 0 setdash 2 copy moveto
	2 copy vpt 0 180 arc closepath fill
	2 copy moveto
	2 copy vpt 270 360 arc closepath fill
	vpt 0 360 arc closepath} bind def
/C12 {BL [] 0 setdash 2 copy moveto
	2 copy vpt 180 360 arc closepath fill
	vpt 0 360 arc closepath} bind def
/C13 {BL [] 0 setdash 2 copy moveto
	2 copy vpt 0 90 arc closepath fill
	2 copy moveto
	2 copy vpt 180 360 arc closepath fill
	vpt 0 360 arc closepath} bind def
/C14 {BL [] 0 setdash 2 copy moveto
	2 copy vpt 90 360 arc closepath fill
	vpt 0 360 arc} bind def
/C15 {BL [] 0 setdash 2 copy vpt 0 360 arc closepath fill
	vpt 0 360 arc closepath} bind def
/Rec {newpath 4 2 roll moveto 1 index 0 rlineto 0 exch rlineto
	neg 0 rlineto closepath} bind def
/Square {dup Rec} bind def
/Bsquare {vpt sub exch vpt sub exch vpt2 Square} bind def
/S0 {BL [] 0 setdash 2 copy moveto 0 vpt rlineto BL Bsquare} bind def
/S1 {BL [] 0 setdash 2 copy vpt Square fill Bsquare} bind def
/S2 {BL [] 0 setdash 2 copy exch vpt sub exch vpt Square fill Bsquare} bind def
/S3 {BL [] 0 setdash 2 copy exch vpt sub exch vpt2 vpt Rec fill Bsquare} bind def
/S4 {BL [] 0 setdash 2 copy exch vpt sub exch vpt sub vpt Square fill Bsquare} bind def
/S5 {BL [] 0 setdash 2 copy 2 copy vpt Square fill
	exch vpt sub exch vpt sub vpt Square fill Bsquare} bind def
/S6 {BL [] 0 setdash 2 copy exch vpt sub exch vpt sub vpt vpt2 Rec fill Bsquare} bind def
/S7 {BL [] 0 setdash 2 copy exch vpt sub exch vpt sub vpt vpt2 Rec fill
	2 copy vpt Square fill Bsquare} bind def
/S8 {BL [] 0 setdash 2 copy vpt sub vpt Square fill Bsquare} bind def
/S9 {BL [] 0 setdash 2 copy vpt sub vpt vpt2 Rec fill Bsquare} bind def
/S10 {BL [] 0 setdash 2 copy vpt sub vpt Square fill 2 copy exch vpt sub exch vpt Square fill
	Bsquare} bind def
/S11 {BL [] 0 setdash 2 copy vpt sub vpt Square fill 2 copy exch vpt sub exch vpt2 vpt Rec fill
	Bsquare} bind def
/S12 {BL [] 0 setdash 2 copy exch vpt sub exch vpt sub vpt2 vpt Rec fill Bsquare} bind def
/S13 {BL [] 0 setdash 2 copy exch vpt sub exch vpt sub vpt2 vpt Rec fill
	2 copy vpt Square fill Bsquare} bind def
/S14 {BL [] 0 setdash 2 copy exch vpt sub exch vpt sub vpt2 vpt Rec fill
	2 copy exch vpt sub exch vpt Square fill Bsquare} bind def
/S15 {BL [] 0 setdash 2 copy Bsquare fill Bsquare} bind def
/D0 {gsave translate 45 rotate 0 0 S0 stroke grestore} bind def
/D1 {gsave translate 45 rotate 0 0 S1 stroke grestore} bind def
/D2 {gsave translate 45 rotate 0 0 S2 stroke grestore} bind def
/D3 {gsave translate 45 rotate 0 0 S3 stroke grestore} bind def
/D4 {gsave translate 45 rotate 0 0 S4 stroke grestore} bind def
/D5 {gsave translate 45 rotate 0 0 S5 stroke grestore} bind def
/D6 {gsave translate 45 rotate 0 0 S6 stroke grestore} bind def
/D7 {gsave translate 45 rotate 0 0 S7 stroke grestore} bind def
/D8 {gsave translate 45 rotate 0 0 S8 stroke grestore} bind def
/D9 {gsave translate 45 rotate 0 0 S9 stroke grestore} bind def
/D10 {gsave translate 45 rotate 0 0 S10 stroke grestore} bind def
/D11 {gsave translate 45 rotate 0 0 S11 stroke grestore} bind def
/D12 {gsave translate 45 rotate 0 0 S12 stroke grestore} bind def
/D13 {gsave translate 45 rotate 0 0 S13 stroke grestore} bind def
/D14 {gsave translate 45 rotate 0 0 S14 stroke grestore} bind def
/D15 {gsave translate 45 rotate 0 0 S15 stroke grestore} bind def
/DiaE {stroke [] 0 setdash vpt add M
  hpt neg vpt neg V hpt vpt neg V
  hpt vpt V hpt neg vpt V closepath stroke} def
/BoxE {stroke [] 0 setdash exch hpt sub exch vpt add M
  0 vpt2 neg V hpt2 0 V 0 vpt2 V
  hpt2 neg 0 V closepath stroke} def
/TriUE {stroke [] 0 setdash vpt 1.12 mul add M
  hpt neg vpt -1.62 mul V
  hpt 2 mul 0 V
  hpt neg vpt 1.62 mul V closepath stroke} def
/TriDE {stroke [] 0 setdash vpt 1.12 mul sub M
  hpt neg vpt 1.62 mul V
  hpt 2 mul 0 V
  hpt neg vpt -1.62 mul V closepath stroke} def
/PentE {stroke [] 0 setdash gsave
  translate 0 hpt M 4 {72 rotate 0 hpt L} repeat
  closepath stroke grestore} def
/CircE {stroke [] 0 setdash 
  hpt 0 360 arc stroke} def
/Opaque {gsave closepath 1 setgray fill grestore 0 setgray closepath} def
/DiaW {stroke [] 0 setdash vpt add M
  hpt neg vpt neg V hpt vpt neg V
  hpt vpt V hpt neg vpt V Opaque stroke} def
/BoxW {stroke [] 0 setdash exch hpt sub exch vpt add M
  0 vpt2 neg V hpt2 0 V 0 vpt2 V
  hpt2 neg 0 V Opaque stroke} def
/TriUW {stroke [] 0 setdash vpt 1.12 mul add M
  hpt neg vpt -1.62 mul V
  hpt 2 mul 0 V
  hpt neg vpt 1.62 mul V Opaque stroke} def
/TriDW {stroke [] 0 setdash vpt 1.12 mul sub M
  hpt neg vpt 1.62 mul V
  hpt 2 mul 0 V
  hpt neg vpt -1.62 mul V Opaque stroke} def
/PentW {stroke [] 0 setdash gsave
  translate 0 hpt M 4 {72 rotate 0 hpt L} repeat
  Opaque stroke grestore} def
/CircW {stroke [] 0 setdash 
  hpt 0 360 arc Opaque stroke} def
/BoxFill {gsave Rec 1 setgray fill grestore} def
/Density {
  /Fillden exch def
  currentrgbcolor
  /ColB exch def /ColG exch def /ColR exch def
  /ColR ColR Fillden mul Fillden sub 1 add def
  /ColG ColG Fillden mul Fillden sub 1 add def
  /ColB ColB Fillden mul Fillden sub 1 add def
  ColR ColG ColB setrgbcolor} def
/BoxColFill {gsave Rec PolyFill} def
/PolyFill {gsave Density fill grestore grestore} def
/h {rlineto rlineto rlineto gsave fill grestore} bind def
%
%
/PatternFill {gsave /PFa [ 9 2 roll ] def
  PFa 0 get PFa 2 get 2 div add PFa 1 get PFa 3 get 2 div add translate
  PFa 2 get -2 div PFa 3 get -2 div PFa 2 get PFa 3 get Rec
  gsave 1 setgray fill grestore clip
  currentlinewidth 0.5 mul setlinewidth
  /PFs PFa 2 get dup mul PFa 3 get dup mul add sqrt def
  0 0 M PFa 5 get rotate PFs -2 div dup translate
  0 1 PFs PFa 4 get div 1 add floor cvi
	{PFa 4 get mul 0 M 0 PFs V} for
  0 PFa 6 get ne {
	0 1 PFs PFa 4 get div 1 add floor cvi
	{PFa 4 get mul 0 2 1 roll M PFs 0 V} for
 } if
  stroke grestore} def
/languagelevel where
 {pop languagelevel} {1} ifelse
 2 lt
	{/InterpretLevel1 true def}
	{/InterpretLevel1 Level1 def}
 ifelse
%
%
/Level2PatternFill {
/Tile8x8 {/PaintType 2 /PatternType 1 /TilingType 1 /BBox [0 0 8 8] /XStep 8 /YStep 8}
	bind def
/KeepColor {currentrgbcolor [/Pattern /DeviceRGB] setcolorspace} bind def
<< Tile8x8
 /PaintProc {0.5 setlinewidth pop 0 0 M 8 8 L 0 8 M 8 0 L stroke} 
>> matrix makepattern
/Pat1 exch def
<< Tile8x8
 /PaintProc {0.5 setlinewidth pop 0 0 M 8 8 L 0 8 M 8 0 L stroke
	0 4 M 4 8 L 8 4 L 4 0 L 0 4 L stroke}
>> matrix makepattern
/Pat2 exch def
<< Tile8x8
 /PaintProc {0.5 setlinewidth pop 0 0 M 0 8 L
	8 8 L 8 0 L 0 0 L fill}
>> matrix makepattern
/Pat3 exch def
<< Tile8x8
 /PaintProc {0.5 setlinewidth pop -4 8 M 8 -4 L
	0 12 M 12 0 L stroke}
>> matrix makepattern
/Pat4 exch def
<< Tile8x8
 /PaintProc {0.5 setlinewidth pop -4 0 M 8 12 L
	0 -4 M 12 8 L stroke}
>> matrix makepattern
/Pat5 exch def
<< Tile8x8
 /PaintProc {0.5 setlinewidth pop -2 8 M 4 -4 L
	0 12 M 8 -4 L 4 12 M 10 0 L stroke}
>> matrix makepattern
/Pat6 exch def
<< Tile8x8
 /PaintProc {0.5 setlinewidth pop -2 0 M 4 12 L
	0 -4 M 8 12 L 4 -4 M 10 8 L stroke}
>> matrix makepattern
/Pat7 exch def
<< Tile8x8
 /PaintProc {0.5 setlinewidth pop 8 -2 M -4 4 L
	12 0 M -4 8 L 12 4 M 0 10 L stroke}
>> matrix makepattern
/Pat8 exch def
<< Tile8x8
 /PaintProc {0.5 setlinewidth pop 0 -2 M 12 4 L
	-4 0 M 12 8 L -4 4 M 8 10 L stroke}
>> matrix makepattern
/Pat9 exch def
/Pattern1 {PatternBgnd KeepColor Pat1 setpattern} bind def
/Pattern2 {PatternBgnd KeepColor Pat2 setpattern} bind def
/Pattern3 {PatternBgnd KeepColor Pat3 setpattern} bind def
/Pattern4 {PatternBgnd KeepColor Landscape {Pat5} {Pat4} ifelse setpattern} bind def
/Pattern5 {PatternBgnd KeepColor Landscape {Pat4} {Pat5} ifelse setpattern} bind def
/Pattern6 {PatternBgnd KeepColor Landscape {Pat9} {Pat6} ifelse setpattern} bind def
/Pattern7 {PatternBgnd KeepColor Landscape {Pat8} {Pat7} ifelse setpattern} bind def
} def
%
%
%
/PatternBgnd {
  TransparentPatterns {} {gsave 1 setgray fill grestore} ifelse
} def
%
%
/Level1PatternFill {
/Pattern1 {0.250 Density} bind def
/Pattern2 {0.500 Density} bind def
/Pattern3 {0.750 Density} bind def
/Pattern4 {0.125 Density} bind def
/Pattern5 {0.375 Density} bind def
/Pattern6 {0.625 Density} bind def
/Pattern7 {0.875 Density} bind def
} def
%
%
Level1 {Level1PatternFill} {Level2PatternFill} ifelse
/Symbol-Oblique /Symbol findfont [1 0 .167 1 0 0] makefont
dup length dict begin {1 index /FID eq {pop pop} {def} ifelse} forall
currentdict end definefont pop
end
gnudict begin
gsave
0 0 translate
0.050 0.050 scale
0 setgray
newpath
1.000 UL
LTb
900 600 M
63 0 V
5897 0 R
-63 0 V
900 1440 M
63 0 V
5897 0 R
-63 0 V
900 2280 M
63 0 V
5897 0 R
-63 0 V
900 3120 M
63 0 V
5897 0 R
-63 0 V
900 3960 M
63 0 V
5897 0 R
-63 0 V
900 4800 M
63 0 V
5897 0 R
-63 0 V
900 600 M
0 63 V
0 4137 R
0 -63 V
2092 600 M
0 63 V
0 4137 R
0 -63 V
3284 600 M
0 63 V
0 4137 R
0 -63 V
4476 600 M
0 63 V
0 4137 R
0 -63 V
5668 600 M
0 63 V
0 4137 R
0 -63 V
6860 600 M
0 63 V
0 4137 R
0 -63 V
900 4800 M
900 600 L
5960 0 V
0 4200 V
-5960 0 V
stroke
LCb setrgbcolor
LTb
LCb setrgbcolor
LTb
1.000 UP
1.000 UL
LTb
0.800 UP
0.500 UL
LT0
1367 982 M
0 3 V
-31 -3 R
62 0 V
-62 3 R
62 0 V
383 197 R
0 3 V
-31 -3 R
62 0 V
-62 3 R
62 0 V
384 176 R
0 5 V
-31 -5 R
62 0 V
-62 5 R
62 0 V
384 156 R
0 7 V
-31 -7 R
62 0 V
-62 7 R
62 0 V
384 152 R
0 7 V
-31 -7 R
62 0 V
-62 7 R
62 0 V
383 149 R
0 6 V
-31 -6 R
62 0 V
-62 6 R
62 0 V
384 152 R
0 8 V
-31 -8 R
62 0 V
-62 8 R
62 0 V
384 147 R
0 9 V
-31 -9 R
62 0 V
-62 9 R
62 0 V
383 138 R
0 10 V
-31 -10 R
62 0 V
-62 10 R
62 0 V
384 136 R
0 9 V
-31 -9 R
62 0 V
-62 9 R
62 0 V
1213 444 R
0 12 V
-31 -12 R
62 0 V
-62 12 R
62 0 V
1367 984 BoxF
1781 1184 BoxF
2196 1363 BoxF
2611 1525 BoxF
3026 1685 BoxF
3440 1840 BoxF
3855 1999 BoxF
4270 2154 BoxF
4684 2302 BoxF
5099 2448 BoxF
6343 2902 BoxF
0.500 UL
LTb
960 700 M
60 71 V
61 51 V
60 43 V
60 38 V
60 36 V
60 33 V
61 32 V
60 30 V
60 30 V
60 28 V
60 28 V
61 27 V
60 27 V
60 26 V
60 26 V
60 26 V
61 25 V
60 25 V
60 25 V
60 24 V
60 25 V
61 24 V
60 24 V
60 24 V
60 23 V
60 24 V
61 23 V
60 24 V
60 23 V
60 23 V
60 23 V
61 23 V
60 23 V
60 23 V
60 23 V
60 23 V
61 22 V
60 23 V
60 22 V
60 23 V
60 22 V
61 23 V
60 22 V
60 23 V
60 22 V
60 22 V
61 22 V
60 23 V
60 22 V
60 22 V
61 22 V
60 22 V
60 22 V
60 22 V
60 22 V
61 22 V
60 22 V
60 22 V
60 22 V
60 22 V
61 22 V
60 22 V
60 22 V
60 22 V
60 22 V
61 21 V
60 22 V
60 22 V
60 22 V
60 21 V
61 22 V
60 22 V
60 22 V
60 21 V
60 22 V
61 22 V
60 22 V
60 21 V
60 22 V
60 22 V
61 21 V
60 22 V
60 21 V
60 22 V
60 22 V
61 21 V
60 22 V
60 22 V
60 21 V
60 22 V
61 21 V
60 22 V
60 21 V
60 22 V
60 21 V
61 22 V
60 22 V
60 21 V
0.800 UP
stroke
LT0
1367 2255 M
0 81 V
-31 -81 R
62 0 V
-62 81 R
62 0 V
383 182 R
0 35 V
-31 -35 R
62 0 V
-62 35 R
62 0 V
384 219 R
0 12 V
-31 -12 R
62 0 V
-62 12 R
62 0 V
384 59 R
0 15 V
-31 -15 R
62 0 V
-62 15 R
62 0 V
384 74 R
0 16 V
-31 -16 R
62 0 V
-62 16 R
62 0 V
383 55 R
0 19 V
-31 -19 R
62 0 V
-62 19 R
62 0 V
384 99 R
0 20 V
-31 -20 R
62 0 V
-62 20 R
62 0 V
384 55 R
0 23 V
-31 -23 R
62 0 V
-62 23 R
62 0 V
383 86 R
0 18 V
-31 -18 R
62 0 V
-62 18 R
62 0 V
384 68 R
0 20 V
-31 -20 R
62 0 V
-62 20 R
62 0 V
1213 303 R
0 28 V
-31 -28 R
62 0 V
-62 28 R
62 0 V
1367 2295 Box
1781 2535 Box
2196 2778 Box
2611 2851 Box
3026 2940 Box
3440 3013 Box
3855 3131 Box
4270 3207 Box
4684 3314 Box
5099 3401 Box
6343 3728 Box
0.500 UL
LTb
900 2704 M
60 4 V
60 4 V
61 5 V
60 5 V
60 5 V
60 6 V
60 5 V
61 6 V
60 6 V
60 6 V
60 6 V
60 7 V
61 6 V
60 7 V
60 7 V
60 8 V
60 7 V
61 8 V
60 7 V
60 8 V
60 9 V
60 8 V
61 8 V
60 9 V
60 9 V
60 9 V
60 9 V
61 9 V
60 10 V
60 9 V
60 10 V
60 10 V
61 10 V
60 10 V
60 10 V
60 11 V
60 11 V
61 10 V
60 11 V
60 11 V
60 11 V
60 12 V
61 11 V
60 12 V
60 12 V
60 11 V
60 12 V
61 12 V
60 13 V
60 12 V
60 12 V
61 13 V
60 13 V
60 12 V
60 13 V
60 13 V
61 13 V
60 13 V
60 14 V
60 13 V
60 14 V
61 13 V
60 14 V
60 14 V
60 14 V
60 14 V
61 14 V
60 14 V
60 14 V
60 15 V
60 14 V
61 15 V
60 14 V
60 15 V
60 15 V
60 15 V
61 14 V
60 15 V
60 16 V
60 15 V
60 15 V
61 15 V
60 16 V
60 15 V
60 16 V
60 15 V
61 16 V
60 16 V
60 15 V
60 16 V
60 16 V
61 16 V
60 16 V
60 16 V
60 16 V
60 17 V
61 16 V
60 16 V
60 17 V
0.800 UP
stroke
LT0
1367 2894 M
0 30 V
-31 -30 R
62 0 V
-62 30 R
62 0 V
-31 621 R
0 62 V
-31 -62 R
62 0 V
-62 62 R
62 0 V
383 -432 R
0 14 V
-31 -14 R
62 0 V
-62 14 R
62 0 V
-31 362 R
0 73 V
-31 -73 R
62 0 V
-62 73 R
62 0 V
384 -274 R
0 18 V
-31 -18 R
62 0 V
-62 18 R
62 0 V
-31 338 R
0 29 V
-31 -29 R
62 0 V
-62 29 R
62 0 V
384 -256 R
0 29 V
-31 -29 R
62 0 V
-62 29 R
62 0 V
-31 202 R
0 33 V
-31 -33 R
62 0 V
-62 33 R
62 0 V
384 -191 R
0 25 V
-31 -25 R
62 0 V
-62 25 R
62 0 V
-31 124 R
0 31 V
-31 -31 R
62 0 V
-62 31 R
62 0 V
383 -56 R
0 25 V
-31 -25 R
62 0 V
-62 25 R
62 0 V
-31 24 R
0 32 V
-31 -32 R
62 0 V
-62 32 R
62 0 V
384 15 R
0 40 V
-31 -40 R
62 0 V
-62 40 R
62 0 V
-31 50 R
0 37 V
-31 -37 R
62 0 V
-62 37 R
62 0 V
384 10 R
0 35 V
-31 -35 R
62 0 V
-62 35 R
62 0 V
-31 -33 R
0 36 V
-31 -36 R
62 0 V
-62 36 R
62 0 V
383 28 R
0 33 V
-31 -33 R
62 0 V
-62 33 R
62 0 V
-31 -25 R
0 39 V
-31 -39 R
4715 3983 L
-62 39 R
62 0 V
384 39 R
0 35 V
-31 -35 R
62 0 V
-62 35 R
62 0 V
-31 -35 R
0 36 V
-31 -36 R
62 0 V
-62 36 R
62 0 V
1213 235 R
0 43 V
-31 -43 R
62 0 V
-62 43 R
62 0 V
-31 -18 R
0 42 V
-31 -42 R
62 0 V
-62 42 R
62 0 V
1367 2909 Circle
1367 3576 Circle
1781 3182 Circle
1781 3587 Circle
2196 3359 Circle
2196 3720 Circle
2611 3494 Circle
2611 3727 Circle
3026 3565 Circle
3026 3716 Circle
3440 3689 Circle
3440 3741 Circle
3855 3792 Circle
3855 3881 Circle
4270 3926 Circle
4270 3929 Circle
4684 3991 Circle
4684 4002 Circle
5099 4079 Circle
5099 4079 Circle
6343 4353 Circle
6343 4378 Circle
0.800 UP
0.500 UL
LT2
1367 3591 M
0 68 V
-31 -68 R
62 0 V
-62 68 R
62 0 V
-31 395 R
0 25 V
-31 -25 R
62 0 V
-62 25 R
62 0 V
383 -405 R
0 22 V
-31 -22 R
62 0 V
-62 22 R
62 0 V
-31 179 R
0 21 V
-31 -21 R
62 0 V
-62 21 R
62 0 V
384 -118 R
0 28 V
-31 -28 R
62 0 V
-62 28 R
62 0 V
-31 129 R
0 33 V
-31 -33 R
62 0 V
-62 33 R
62 0 V
384 -181 R
0 34 V
-31 -34 R
62 0 V
-62 34 R
62 0 V
-31 140 R
0 44 V
-31 -44 R
62 0 V
-62 44 R
62 0 V
384 -201 R
0 41 V
-31 -41 R
62 0 V
-62 41 R
62 0 V
-31 29 R
0 31 V
-31 -31 R
62 0 V
-62 31 R
62 0 V
383 -46 R
0 41 V
-31 -41 R
62 0 V
-62 41 R
62 0 V
-31 -51 R
0 36 V
-31 -36 R
62 0 V
-62 36 R
62 0 V
384 1 R
0 36 V
-31 -36 R
62 0 V
-62 36 R
62 0 V
-31 4 R
0 39 V
-31 -39 R
62 0 V
-62 39 R
62 0 V
384 6 R
0 41 V
-31 -41 R
62 0 V
-62 41 R
62 0 V
-31 -19 R
0 40 V
-31 -40 R
62 0 V
-62 40 R
62 0 V
383 36 R
0 41 V
-31 -41 R
62 0 V
-62 41 R
62 0 V
-31 -29 R
0 38 V
-31 -38 R
4715 4081 L
-62 38 R
62 0 V
384 -4 R
0 47 V
-31 -47 R
62 0 V
-62 47 R
62 0 V
-31 -25 R
0 49 V
-31 -49 R
62 0 V
-62 49 R
62 0 V
1213 187 R
0 37 V
-31 -37 R
62 0 V
-62 37 R
62 0 V
-31 -13 R
0 34 V
-31 -34 R
62 0 V
-62 34 R
62 0 V
1367 3625 TriU
1367 4067 TriU
1781 3685 TriU
1781 3885 TriU
2196 3792 TriU
2196 3952 TriU
2611 3804 TriU
2611 3983 TriU
3026 3825 TriU
3026 3890 TriU
3440 3879 TriU
3440 3867 TriU
3855 3904 TriU
3855 3946 TriU
4270 3991 TriU
4270 4013 TriU
4684 4090 TriU
4684 4100 TriU
5099 4138 TriU
5099 4161 TriU
6343 4392 TriU
6343 4414 TriU
0.500 UL
LTb
900 3576 M
60 3 V
60 4 V
61 3 V
60 3 V
60 4 V
60 4 V
60 4 V
61 4 V
60 4 V
60 5 V
60 4 V
60 5 V
61 5 V
60 5 V
60 5 V
60 5 V
60 5 V
61 6 V
60 5 V
60 6 V
60 6 V
60 6 V
61 6 V
60 7 V
60 6 V
60 7 V
60 6 V
61 7 V
60 7 V
60 7 V
60 7 V
60 8 V
61 7 V
60 8 V
60 8 V
60 7 V
60 8 V
61 8 V
60 9 V
60 8 V
60 8 V
60 9 V
61 9 V
60 8 V
60 9 V
60 9 V
60 10 V
61 9 V
60 9 V
60 10 V
60 9 V
61 10 V
60 10 V
60 10 V
60 10 V
60 10 V
61 10 V
60 10 V
60 11 V
60 10 V
60 11 V
61 10 V
60 11 V
60 11 V
60 11 V
60 11 V
61 12 V
60 11 V
60 11 V
60 12 V
60 11 V
61 12 V
60 12 V
60 11 V
60 12 V
60 12 V
61 12 V
60 13 V
60 12 V
60 12 V
60 13 V
61 12 V
60 13 V
60 12 V
60 13 V
60 13 V
61 13 V
60 13 V
60 13 V
60 13 V
60 13 V
61 13 V
60 14 V
60 13 V
60 14 V
60 13 V
61 14 V
60 13 V
60 14 V
0.800 UP
stroke
LT4
2198 1359 M
0 7 V
-31 -7 R
62 0 V
-62 7 R
62 0 V
799 323 R
0 8 V
-31 -8 R
62 0 V
-62 8 R
62 0 V
384 148 R
0 12 V
-31 -12 R
62 0 V
-62 12 R
62 0 V
1214 437 R
0 7 V
-31 -7 R
62 0 V
-62 7 R
62 0 V
2198 1363 Pls
3028 1693 Pls
3443 1851 Pls
4688 2297 Pls
0.800 UP
0.500 UL
LT4
2198 2804 M
0 8 V
-31 -8 R
62 0 V
-62 8 R
62 0 V
799 142 R
0 10 V
-31 -10 R
62 0 V
-62 10 R
62 0 V
384 52 R
0 18 V
-31 -18 R
62 0 V
-62 18 R
62 0 V
1214 277 R
0 9 V
-31 -9 R
62 0 V
-62 9 R
62 0 V
2198 2808 Crs
3028 2959 Crs
3443 3025 Crs
4688 3316 Crs
0.800 UP
0.500 UL
LT4
2198 3255 M
0 49 V
-31 -49 R
62 0 V
-62 49 R
62 0 V
-31 388 R
0 22 V
-31 -22 R
62 0 V
-62 22 R
62 0 V
799 -38 R
0 14 V
-31 -14 R
62 0 V
-62 14 R
62 0 V
-31 43 R
0 13 V
-31 -13 R
62 0 V
-62 13 R
62 0 V
384 -45 R
0 18 V
-31 -18 R
62 0 V
-62 18 R
62 0 V
-31 5 R
0 27 V
-31 -27 R
62 0 V
-62 27 R
62 0 V
1214 279 R
0 26 V
-31 -26 R
62 0 V
-62 26 R
62 0 V
-31 -123 R
0 59 V
-31 -59 R
62 0 V
-62 59 R
62 0 V
2198 3280 Star
2198 3703 Star
3028 3683 Star
3028 3740 Star
3443 3710 Star
3443 3737 Star
4688 4043 Star
4688 3963 Star
0.800 UP
0.500 UL
LT4
2198 2804 M
0 8 V
-31 -8 R
62 0 V
-62 8 R
62 0 V
799 142 R
0 10 V
-31 -10 R
62 0 V
-62 10 R
62 0 V
384 52 R
0 18 V
-31 -18 R
62 0 V
-62 18 R
62 0 V
1214 277 R
0 9 V
-31 -9 R
62 0 V
-62 9 R
62 0 V
2198 2808 Crs
3028 2959 Crs
3443 3025 Crs
4688 3316 Crs
0.800 UP
0.500 UL
LT4
2198 3255 M
0 49 V
-31 -49 R
62 0 V
-62 49 R
62 0 V
-31 388 R
0 22 V
-31 -22 R
62 0 V
-62 22 R
62 0 V
799 -38 R
0 14 V
-31 -14 R
62 0 V
-62 14 R
62 0 V
-31 43 R
0 13 V
-31 -13 R
62 0 V
-62 13 R
62 0 V
384 -45 R
0 18 V
-31 -18 R
62 0 V
-62 18 R
62 0 V
-31 5 R
0 27 V
-31 -27 R
62 0 V
-62 27 R
62 0 V
1214 279 R
0 26 V
-31 -26 R
62 0 V
-62 26 R
62 0 V
-31 -123 R
0 59 V
-31 -59 R
62 0 V
-62 59 R
62 0 V
2198 3280 Star
2198 3703 Star
3028 3683 Star
3028 3740 Star
3443 3710 Star
3443 3737 Star
4688 4043 Star
4688 3963 Star
0.800 UP
0.500 UL
LT3
2198 3791 M
0 22 V
-31 -22 R
62 0 V
-62 22 R
62 0 V
-31 52 R
0 46 V
-31 -46 R
62 0 V
-62 46 R
62 0 V
799 -94 R
0 27 V
-31 -27 R
62 0 V
-62 27 R
62 0 V
-31 103 R
0 30 V
-31 -30 R
62 0 V
-62 30 R
62 0 V
384 -101 R
0 17 V
-31 -17 R
62 0 V
-62 17 R
62 0 V
-31 -14 R
0 33 V
-31 -33 R
62 0 V
-62 33 R
62 0 V
1214 158 R
0 38 V
-31 -38 R
62 0 V
-62 38 R
62 0 V
-31 -71 R
0 76 V
-31 -76 R
62 0 V
-62 76 R
62 0 V
2198 3802 Dia
2198 3888 Dia
3028 3831 Dia
3028 3962 Dia
3443 3884 Dia
3443 3896 Dia
4688 4089 Dia
4688 4075 Dia
1.500 UL
LT1
3669 3518 M
60 -2 V
61 -1 V
60 0 V
60 0 V
60 1 V
61 2 V
60 2 V
60 2 V
60 4 V
60 3 V
61 4 V
60 5 V
60 4 V
60 6 V
60 5 V
61 6 V
60 7 V
60 7 V
60 7 V
60 7 V
61 8 V
60 8 V
60 8 V
60 9 V
60 9 V
61 9 V
60 10 V
60 9 V
60 10 V
60 11 V
61 10 V
60 11 V
60 10 V
60 11 V
60 12 V
61 11 V
60 12 V
60 12 V
60 12 V
60 12 V
61 12 V
60 13 V
60 12 V
60 13 V
60 13 V
61 13 V
60 14 V
60 13 V
60 14 V
60 13 V
61 14 V
60 14 V
60 14 V
3669 2688 M
60 35 V
61 34 V
60 32 V
60 31 V
60 30 V
61 29 V
60 28 V
60 27 V
60 27 V
60 25 V
61 25 V
60 25 V
60 23 V
60 23 V
60 23 V
61 23 V
60 21 V
60 22 V
60 21 V
60 21 V
61 20 V
60 21 V
60 20 V
60 20 V
60 19 V
61 20 V
60 19 V
60 19 V
60 19 V
60 19 V
61 18 V
60 19 V
60 18 V
60 19 V
60 18 V
61 18 V
60 18 V
60 18 V
60 18 V
60 18 V
61 18 V
60 18 V
60 18 V
60 18 V
60 17 V
61 18 V
60 18 V
60 18 V
60 17 V
60 18 V
6740 3787 L
60 17 V
60 18 V
stroke
1.000 UL
LTb
900 4800 M
900 600 L
5960 0 V
0 4200 V
-5960 0 V
1.000 UP
stroke
grestore
end
showpage
  }}%
  \put(3880,100){\makebox(0,0){\strut{}$\sqrt{\sigma} l$}}%
  \put(200,2700){%
  \special{ps: gsave currentpoint currentpoint translate
 0 rotate neg exch neg exch translate}%
  \makebox(0,0){\strut{}$E / \sqrt{\sigma}$}%
  \special{ps: currentpoint grestore moveto}%
  }%
  \put(6860,400){\makebox(0,0){\strut{} 6}}%
  \put(5668,400){\makebox(0,0){\strut{} 5}}%
  \put(4476,400){\makebox(0,0){\strut{} 4}}%
  \put(3284,400){\makebox(0,0){\strut{} 3}}%
  \put(2092,400){\makebox(0,0){\strut{} 2}}%
  \put(900,400){\makebox(0,0){\strut{} 1}}%
  \put(780,4800){\makebox(0,0)[r]{\strut{} 10}}%
  \put(780,3960){\makebox(0,0)[r]{\strut{} 8}}%
  \put(780,3120){\makebox(0,0)[r]{\strut{} 6}}%
  \put(780,2280){\makebox(0,0)[r]{\strut{} 4}}%
  \put(780,1440){\makebox(0,0)[r]{\strut{} 2}}%
  \put(780,600){\makebox(0,0)[r]{\strut{} 0}}%
\end{picture}%
\endgroup
 

%% file: su6all2.tex
\begingroup%
\makeatletter%
\newcommand{\GNUPLOTspecial}{%
  \@sanitize\catcode`\%=14\relax\special}%
\setlength{\unitlength}{0.0500bp}%
\begin{picture}(7200,5040)(0,0)%
  {\GNUPLOTspecial{"
/gnudict 256 dict def
gnudict begin
%
%
/Color true def
/Blacktext false def
/Solid true def
/Dashlength 1 def
/Landscape false def
/Level1 false def
/Rounded false def
/TransparentPatterns false def
/gnulinewidth 5.000 def
/userlinewidth gnulinewidth def
/vshift -66 def
/dl1 {
  10.0 Dashlength mul mul
  Rounded { currentlinewidth 0.75 mul sub dup 0 le { pop 0.01 } if } if
} def
/dl2 {
  10.0 Dashlength mul mul
  Rounded { currentlinewidth 0.75 mul add } if
} def
/hpt_ 31.5 def
/vpt_ 31.5 def
/hpt hpt_ def
/vpt vpt_ def
Level1 {} {
/SDict 10 dict def
systemdict /pdfmark known not {
  userdict /pdfmark systemdict /cleartomark get put
} if
SDict begin [
  /Title (../../../../../../../LATTICE/POS/PLOTS/su6all2.tex)
  /Subject (gnuplot plot)
  /Creator (gnuplot 4.2 patchlevel 0)
  /Author (U-YOUR-4F60B1CEA3\\Andreas,S-1-5-21-2307956900-2123931569-774947135-1006)
  /CreationDate (Mon Sep 10 14:24:58 2007)
  /DOCINFO pdfmark
end
} ifelse
%
%
/M {moveto} bind def
/L {lineto} bind def
/R {rmoveto} bind def
/V {rlineto} bind def
/N {newpath moveto} bind def
/Z {closepath} bind def
/C {setrgbcolor} bind def
/f {rlineto fill} bind def
/vpt2 vpt 2 mul def
/hpt2 hpt 2 mul def
/Lshow {currentpoint stroke M 0 vshift R 
	Blacktext {gsave 0 setgray show grestore} {show} ifelse} def
/Rshow {currentpoint stroke M dup stringwidth pop neg vshift R
	Blacktext {gsave 0 setgray show grestore} {show} ifelse} def
/Cshow {currentpoint stroke M dup stringwidth pop -2 div vshift R 
	Blacktext {gsave 0 setgray show grestore} {show} ifelse} def
/UP {dup vpt_ mul /vpt exch def hpt_ mul /hpt exch def
  /hpt2 hpt 2 mul def /vpt2 vpt 2 mul def} def
/DL {Color {setrgbcolor Solid {pop []} if 0 setdash}
 {pop pop pop 0 setgray Solid {pop []} if 0 setdash} ifelse} def
/BL {stroke userlinewidth 2 mul setlinewidth
	Rounded {1 setlinejoin 1 setlinecap} if} def
/AL {stroke userlinewidth 2 div setlinewidth
	Rounded {1 setlinejoin 1 setlinecap} if} def
/UL {dup gnulinewidth mul /userlinewidth exch def
	dup 1 lt {pop 1} if 10 mul /udl exch def} def
/PL {stroke userlinewidth setlinewidth
	Rounded {1 setlinejoin 1 setlinecap} if} def
/LCw {1 1 1} def
/LCb {0 0 0} def
/LCa {0 0 0} def
/LC0 {1 0 0} def
/LC1 {0 1 0} def
/LC2 {0 0 1} def
/LC3 {1 0 1} def
/LC4 {0 1 1} def
/LC5 {1 1 0} def
/LC6 {0 0 0} def
/LC7 {1 0.3 0} def
/LC8 {0.5 0.5 0.5} def
/LTw {PL [] 1 setgray} def
/LTb {BL [] LCb DL} def
/LTa {AL [1 udl mul 2 udl mul] 0 setdash LCa setrgbcolor} def
/LT0 {PL [] LC0 DL} def
/LT1 {PL [4 dl1 2 dl2] LC1 DL} def
/LT2 {PL [2 dl1 3 dl2] LC2 DL} def
/LT3 {PL [1 dl1 1.5 dl2] LC3 DL} def
/LT4 {PL [6 dl1 2 dl2 1 dl1 2 dl2] LC4 DL} def
/LT5 {PL [3 dl1 3 dl2 1 dl1 3 dl2] LC5 DL} def
/LT6 {PL [2 dl1 2 dl2 2 dl1 6 dl2] LC6 DL} def
/LT7 {PL [1 dl1 2 dl2 6 dl1 2 dl2 1 dl1 2 dl2] LC7 DL} def
/LT8 {PL [2 dl1 2 dl2 2 dl1 2 dl2 2 dl1 2 dl2 2 dl1 4 dl2] LC8 DL} def
/Pnt {stroke [] 0 setdash gsave 1 setlinecap M 0 0 V stroke grestore} def
/Dia {stroke [] 0 setdash 2 copy vpt add M
  hpt neg vpt neg V hpt vpt neg V
  hpt vpt V hpt neg vpt V closepath stroke
  Pnt} def
/Pls {stroke [] 0 setdash vpt sub M 0 vpt2 V
  currentpoint stroke M
  hpt neg vpt neg R hpt2 0 V stroke
 } def
/Box {stroke [] 0 setdash 2 copy exch hpt sub exch vpt add M
  0 vpt2 neg V hpt2 0 V 0 vpt2 V
  hpt2 neg 0 V closepath stroke
  Pnt} def
/Crs {stroke [] 0 setdash exch hpt sub exch vpt add M
  hpt2 vpt2 neg V currentpoint stroke M
  hpt2 neg 0 R hpt2 vpt2 V stroke} def
/TriU {stroke [] 0 setdash 2 copy vpt 1.12 mul add M
  hpt neg vpt -1.62 mul V
  hpt 2 mul 0 V
  hpt neg vpt 1.62 mul V closepath stroke
  Pnt} def
/Star {2 copy Pls Crs} def
/BoxF {stroke [] 0 setdash exch hpt sub exch vpt add M
  0 vpt2 neg V hpt2 0 V 0 vpt2 V
  hpt2 neg 0 V closepath fill} def
/TriUF {stroke [] 0 setdash vpt 1.12 mul add M
  hpt neg vpt -1.62 mul V
  hpt 2 mul 0 V
  hpt neg vpt 1.62 mul V closepath fill} def
/TriD {stroke [] 0 setdash 2 copy vpt 1.12 mul sub M
  hpt neg vpt 1.62 mul V
  hpt 2 mul 0 V
  hpt neg vpt -1.62 mul V closepath stroke
  Pnt} def
/TriDF {stroke [] 0 setdash vpt 1.12 mul sub M
  hpt neg vpt 1.62 mul V
  hpt 2 mul 0 V
  hpt neg vpt -1.62 mul V closepath fill} def
/DiaF {stroke [] 0 setdash vpt add M
  hpt neg vpt neg V hpt vpt neg V
  hpt vpt V hpt neg vpt V closepath fill} def
/Pent {stroke [] 0 setdash 2 copy gsave
  translate 0 hpt M 4 {72 rotate 0 hpt L} repeat
  closepath stroke grestore Pnt} def
/PentF {stroke [] 0 setdash gsave
  translate 0 hpt M 4 {72 rotate 0 hpt L} repeat
  closepath fill grestore} def
/Circle {stroke [] 0 setdash 2 copy
  hpt 0 360 arc stroke Pnt} def
/CircleF {stroke [] 0 setdash hpt 0 360 arc fill} def
/C0 {BL [] 0 setdash 2 copy moveto vpt 90 450 arc} bind def
/C1 {BL [] 0 setdash 2 copy moveto
	2 copy vpt 0 90 arc closepath fill
	vpt 0 360 arc closepath} bind def
/C2 {BL [] 0 setdash 2 copy moveto
	2 copy vpt 90 180 arc closepath fill
	vpt 0 360 arc closepath} bind def
/C3 {BL [] 0 setdash 2 copy moveto
	2 copy vpt 0 180 arc closepath fill
	vpt 0 360 arc closepath} bind def
/C4 {BL [] 0 setdash 2 copy moveto
	2 copy vpt 180 270 arc closepath fill
	vpt 0 360 arc closepath} bind def
/C5 {BL [] 0 setdash 2 copy moveto
	2 copy vpt 0 90 arc
	2 copy moveto
	2 copy vpt 180 270 arc closepath fill
	vpt 0 360 arc} bind def
/C6 {BL [] 0 setdash 2 copy moveto
	2 copy vpt 90 270 arc closepath fill
	vpt 0 360 arc closepath} bind def
/C7 {BL [] 0 setdash 2 copy moveto
	2 copy vpt 0 270 arc closepath fill
	vpt 0 360 arc closepath} bind def
/C8 {BL [] 0 setdash 2 copy moveto
	2 copy vpt 270 360 arc closepath fill
	vpt 0 360 arc closepath} bind def
/C9 {BL [] 0 setdash 2 copy moveto
	2 copy vpt 270 450 arc closepath fill
	vpt 0 360 arc closepath} bind def
/C10 {BL [] 0 setdash 2 copy 2 copy moveto vpt 270 360 arc closepath fill
	2 copy moveto
	2 copy vpt 90 180 arc closepath fill
	vpt 0 360 arc closepath} bind def
/C11 {BL [] 0 setdash 2 copy moveto
	2 copy vpt 0 180 arc closepath fill
	2 copy moveto
	2 copy vpt 270 360 arc closepath fill
	vpt 0 360 arc closepath} bind def
/C12 {BL [] 0 setdash 2 copy moveto
	2 copy vpt 180 360 arc closepath fill
	vpt 0 360 arc closepath} bind def
/C13 {BL [] 0 setdash 2 copy moveto
	2 copy vpt 0 90 arc closepath fill
	2 copy moveto
	2 copy vpt 180 360 arc closepath fill
	vpt 0 360 arc closepath} bind def
/C14 {BL [] 0 setdash 2 copy moveto
	2 copy vpt 90 360 arc closepath fill
	vpt 0 360 arc} bind def
/C15 {BL [] 0 setdash 2 copy vpt 0 360 arc closepath fill
	vpt 0 360 arc closepath} bind def
/Rec {newpath 4 2 roll moveto 1 index 0 rlineto 0 exch rlineto
	neg 0 rlineto closepath} bind def
/Square {dup Rec} bind def
/Bsquare {vpt sub exch vpt sub exch vpt2 Square} bind def
/S0 {BL [] 0 setdash 2 copy moveto 0 vpt rlineto BL Bsquare} bind def
/S1 {BL [] 0 setdash 2 copy vpt Square fill Bsquare} bind def
/S2 {BL [] 0 setdash 2 copy exch vpt sub exch vpt Square fill Bsquare} bind def
/S3 {BL [] 0 setdash 2 copy exch vpt sub exch vpt2 vpt Rec fill Bsquare} bind def
/S4 {BL [] 0 setdash 2 copy exch vpt sub exch vpt sub vpt Square fill Bsquare} bind def
/S5 {BL [] 0 setdash 2 copy 2 copy vpt Square fill
	exch vpt sub exch vpt sub vpt Square fill Bsquare} bind def
/S6 {BL [] 0 setdash 2 copy exch vpt sub exch vpt sub vpt vpt2 Rec fill Bsquare} bind def
/S7 {BL [] 0 setdash 2 copy exch vpt sub exch vpt sub vpt vpt2 Rec fill
	2 copy vpt Square fill Bsquare} bind def
/S8 {BL [] 0 setdash 2 copy vpt sub vpt Square fill Bsquare} bind def
/S9 {BL [] 0 setdash 2 copy vpt sub vpt vpt2 Rec fill Bsquare} bind def
/S10 {BL [] 0 setdash 2 copy vpt sub vpt Square fill 2 copy exch vpt sub exch vpt Square fill
	Bsquare} bind def
/S11 {BL [] 0 setdash 2 copy vpt sub vpt Square fill 2 copy exch vpt sub exch vpt2 vpt Rec fill
	Bsquare} bind def
/S12 {BL [] 0 setdash 2 copy exch vpt sub exch vpt sub vpt2 vpt Rec fill Bsquare} bind def
/S13 {BL [] 0 setdash 2 copy exch vpt sub exch vpt sub vpt2 vpt Rec fill
	2 copy vpt Square fill Bsquare} bind def
/S14 {BL [] 0 setdash 2 copy exch vpt sub exch vpt sub vpt2 vpt Rec fill
	2 copy exch vpt sub exch vpt Square fill Bsquare} bind def
/S15 {BL [] 0 setdash 2 copy Bsquare fill Bsquare} bind def
/D0 {gsave translate 45 rotate 0 0 S0 stroke grestore} bind def
/D1 {gsave translate 45 rotate 0 0 S1 stroke grestore} bind def
/D2 {gsave translate 45 rotate 0 0 S2 stroke grestore} bind def
/D3 {gsave translate 45 rotate 0 0 S3 stroke grestore} bind def
/D4 {gsave translate 45 rotate 0 0 S4 stroke grestore} bind def
/D5 {gsave translate 45 rotate 0 0 S5 stroke grestore} bind def
/D6 {gsave translate 45 rotate 0 0 S6 stroke grestore} bind def
/D7 {gsave translate 45 rotate 0 0 S7 stroke grestore} bind def
/D8 {gsave translate 45 rotate 0 0 S8 stroke grestore} bind def
/D9 {gsave translate 45 rotate 0 0 S9 stroke grestore} bind def
/D10 {gsave translate 45 rotate 0 0 S10 stroke grestore} bind def
/D11 {gsave translate 45 rotate 0 0 S11 stroke grestore} bind def
/D12 {gsave translate 45 rotate 0 0 S12 stroke grestore} bind def
/D13 {gsave translate 45 rotate 0 0 S13 stroke grestore} bind def
/D14 {gsave translate 45 rotate 0 0 S14 stroke grestore} bind def
/D15 {gsave translate 45 rotate 0 0 S15 stroke grestore} bind def
/DiaE {stroke [] 0 setdash vpt add M
  hpt neg vpt neg V hpt vpt neg V
  hpt vpt V hpt neg vpt V closepath stroke} def
/BoxE {stroke [] 0 setdash exch hpt sub exch vpt add M
  0 vpt2 neg V hpt2 0 V 0 vpt2 V
  hpt2 neg 0 V closepath stroke} def
/TriUE {stroke [] 0 setdash vpt 1.12 mul add M
  hpt neg vpt -1.62 mul V
  hpt 2 mul 0 V
  hpt neg vpt 1.62 mul V closepath stroke} def
/TriDE {stroke [] 0 setdash vpt 1.12 mul sub M
  hpt neg vpt 1.62 mul V
  hpt 2 mul 0 V
  hpt neg vpt -1.62 mul V closepath stroke} def
/PentE {stroke [] 0 setdash gsave
  translate 0 hpt M 4 {72 rotate 0 hpt L} repeat
  closepath stroke grestore} def
/CircE {stroke [] 0 setdash 
  hpt 0 360 arc stroke} def
/Opaque {gsave closepath 1 setgray fill grestore 0 setgray closepath} def
/DiaW {stroke [] 0 setdash vpt add M
  hpt neg vpt neg V hpt vpt neg V
  hpt vpt V hpt neg vpt V Opaque stroke} def
/BoxW {stroke [] 0 setdash exch hpt sub exch vpt add M
  0 vpt2 neg V hpt2 0 V 0 vpt2 V
  hpt2 neg 0 V Opaque stroke} def
/TriUW {stroke [] 0 setdash vpt 1.12 mul add M
  hpt neg vpt -1.62 mul V
  hpt 2 mul 0 V
  hpt neg vpt 1.62 mul V Opaque stroke} def
/TriDW {stroke [] 0 setdash vpt 1.12 mul sub M
  hpt neg vpt 1.62 mul V
  hpt 2 mul 0 V
  hpt neg vpt -1.62 mul V Opaque stroke} def
/PentW {stroke [] 0 setdash gsave
  translate 0 hpt M 4 {72 rotate 0 hpt L} repeat
  Opaque stroke grestore} def
/CircW {stroke [] 0 setdash 
  hpt 0 360 arc Opaque stroke} def
/BoxFill {gsave Rec 1 setgray fill grestore} def
/Density {
  /Fillden exch def
  currentrgbcolor
  /ColB exch def /ColG exch def /ColR exch def
  /ColR ColR Fillden mul Fillden sub 1 add def
  /ColG ColG Fillden mul Fillden sub 1 add def
  /ColB ColB Fillden mul Fillden sub 1 add def
  ColR ColG ColB setrgbcolor} def
/BoxColFill {gsave Rec PolyFill} def
/PolyFill {gsave Density fill grestore grestore} def
/h {rlineto rlineto rlineto gsave fill grestore} bind def
%
%
/PatternFill {gsave /PFa [ 9 2 roll ] def
  PFa 0 get PFa 2 get 2 div add PFa 1 get PFa 3 get 2 div add translate
  PFa 2 get -2 div PFa 3 get -2 div PFa 2 get PFa 3 get Rec
  gsave 1 setgray fill grestore clip
  currentlinewidth 0.5 mul setlinewidth
  /PFs PFa 2 get dup mul PFa 3 get dup mul add sqrt def
  0 0 M PFa 5 get rotate PFs -2 div dup translate
  0 1 PFs PFa 4 get div 1 add floor cvi
	{PFa 4 get mul 0 M 0 PFs V} for
  0 PFa 6 get ne {
	0 1 PFs PFa 4 get div 1 add floor cvi
	{PFa 4 get mul 0 2 1 roll M PFs 0 V} for
 } if
  stroke grestore} def
/languagelevel where
 {pop languagelevel} {1} ifelse
 2 lt
	{/InterpretLevel1 true def}
	{/InterpretLevel1 Level1 def}
 ifelse
%
%
/Level2PatternFill {
/Tile8x8 {/PaintType 2 /PatternType 1 /TilingType 1 /BBox [0 0 8 8] /XStep 8 /YStep 8}
	bind def
/KeepColor {currentrgbcolor [/Pattern /DeviceRGB] setcolorspace} bind def
<< Tile8x8
 /PaintProc {0.5 setlinewidth pop 0 0 M 8 8 L 0 8 M 8 0 L stroke} 
>> matrix makepattern
/Pat1 exch def
<< Tile8x8
 /PaintProc {0.5 setlinewidth pop 0 0 M 8 8 L 0 8 M 8 0 L stroke
	0 4 M 4 8 L 8 4 L 4 0 L 0 4 L stroke}
>> matrix makepattern
/Pat2 exch def
<< Tile8x8
 /PaintProc {0.5 setlinewidth pop 0 0 M 0 8 L
	8 8 L 8 0 L 0 0 L fill}
>> matrix makepattern
/Pat3 exch def
<< Tile8x8
 /PaintProc {0.5 setlinewidth pop -4 8 M 8 -4 L
	0 12 M 12 0 L stroke}
>> matrix makepattern
/Pat4 exch def
<< Tile8x8
 /PaintProc {0.5 setlinewidth pop -4 0 M 8 12 L
	0 -4 M 12 8 L stroke}
>> matrix makepattern
/Pat5 exch def
<< Tile8x8
 /PaintProc {0.5 setlinewidth pop -2 8 M 4 -4 L
	0 12 M 8 -4 L 4 12 M 10 0 L stroke}
>> matrix makepattern
/Pat6 exch def
<< Tile8x8
 /PaintProc {0.5 setlinewidth pop -2 0 M 4 12 L
	0 -4 M 8 12 L 4 -4 M 10 8 L stroke}
>> matrix makepattern
/Pat7 exch def
<< Tile8x8
 /PaintProc {0.5 setlinewidth pop 8 -2 M -4 4 L
	12 0 M -4 8 L 12 4 M 0 10 L stroke}
>> matrix makepattern
/Pat8 exch def
<< Tile8x8
 /PaintProc {0.5 setlinewidth pop 0 -2 M 12 4 L
	-4 0 M 12 8 L -4 4 M 8 10 L stroke}
>> matrix makepattern
/Pat9 exch def
/Pattern1 {PatternBgnd KeepColor Pat1 setpattern} bind def
/Pattern2 {PatternBgnd KeepColor Pat2 setpattern} bind def
/Pattern3 {PatternBgnd KeepColor Pat3 setpattern} bind def
/Pattern4 {PatternBgnd KeepColor Landscape {Pat5} {Pat4} ifelse setpattern} bind def
/Pattern5 {PatternBgnd KeepColor Landscape {Pat4} {Pat5} ifelse setpattern} bind def
/Pattern6 {PatternBgnd KeepColor Landscape {Pat9} {Pat6} ifelse setpattern} bind def
/Pattern7 {PatternBgnd KeepColor Landscape {Pat8} {Pat7} ifelse setpattern} bind def
} def
%
%
%
/PatternBgnd {
  TransparentPatterns {} {gsave 1 setgray fill grestore} ifelse
} def
%
%
/Level1PatternFill {
/Pattern1 {0.250 Density} bind def
/Pattern2 {0.500 Density} bind def
/Pattern3 {0.750 Density} bind def
/Pattern4 {0.125 Density} bind def
/Pattern5 {0.375 Density} bind def
/Pattern6 {0.625 Density} bind def
/Pattern7 {0.875 Density} bind def
} def
%
%
Level1 {Level1PatternFill} {Level2PatternFill} ifelse
/Symbol-Oblique /Symbol findfont [1 0 .167 1 0 0] makefont
dup length dict begin {1 index /FID eq {pop pop} {def} ifelse} forall
currentdict end definefont pop
end
gnudict begin
gsave
0 0 translate
0.050 0.050 scale
0 setgray
newpath
1.000 UL
LTb
900 600 M
63 0 V
5897 0 R
-63 0 V
900 1440 M
63 0 V
5897 0 R
-63 0 V
900 2280 M
63 0 V
5897 0 R
-63 0 V
900 3120 M
63 0 V
5897 0 R
-63 0 V
900 3960 M
63 0 V
5897 0 R
-63 0 V
900 4800 M
63 0 V
5897 0 R
-63 0 V
900 600 M
0 63 V
0 4137 R
0 -63 V
2092 600 M
0 63 V
0 4137 R
0 -63 V
3284 600 M
0 63 V
0 4137 R
0 -63 V
4476 600 M
0 63 V
0 4137 R
0 -63 V
5668 600 M
0 63 V
0 4137 R
0 -63 V
6860 600 M
0 63 V
0 4137 R
0 -63 V
900 4800 M
900 600 L
5960 0 V
0 4200 V
-5960 0 V
stroke
LCb setrgbcolor
LTb
LCb setrgbcolor
LTb
1.000 UP
1.000 UL
LTb
0.800 UP
0.500 UL
LT0
2167 1349 M
0 5 V
-31 -5 R
62 0 V
-62 5 R
62 0 V
379 157 R
0 6 V
-31 -6 R
62 0 V
-62 6 R
62 0 V
379 154 R
0 6 V
-31 -6 R
62 0 V
-62 6 R
62 0 V
379 146 R
0 7 V
-31 -7 R
62 0 V
-62 7 R
62 0 V
379 148 R
0 8 V
-31 -8 R
62 0 V
-62 8 R
62 0 V
379 136 R
0 7 V
-31 -7 R
62 0 V
-62 7 R
62 0 V
378 145 R
0 9 V
-31 -9 R
62 0 V
-62 9 R
62 0 V
2167 1352 BoxF
2577 1514 BoxF
2987 1674 BoxF
3397 1827 BoxF
3807 1982 BoxF
4217 2126 BoxF
4626 2279 BoxF
0.500 UL
LTb
960 700 M
60 71 V
61 51 V
60 43 V
60 38 V
60 36 V
60 33 V
61 32 V
60 30 V
60 30 V
60 28 V
60 28 V
61 27 V
60 27 V
60 26 V
60 26 V
60 26 V
61 25 V
60 25 V
60 25 V
60 24 V
60 25 V
61 24 V
60 24 V
60 24 V
60 23 V
60 24 V
61 23 V
60 24 V
60 23 V
60 23 V
60 23 V
61 23 V
60 23 V
60 23 V
60 23 V
60 23 V
61 22 V
60 23 V
60 22 V
60 23 V
60 22 V
61 23 V
60 22 V
60 23 V
60 22 V
60 22 V
61 22 V
60 23 V
60 22 V
60 22 V
61 22 V
60 22 V
60 22 V
60 22 V
60 22 V
61 22 V
60 22 V
60 22 V
60 22 V
60 22 V
61 22 V
60 22 V
60 22 V
60 22 V
60 22 V
61 21 V
60 22 V
60 22 V
60 22 V
60 21 V
61 22 V
60 22 V
60 22 V
60 21 V
60 22 V
61 22 V
60 22 V
60 21 V
60 22 V
60 22 V
61 21 V
60 22 V
60 21 V
60 22 V
60 22 V
61 21 V
60 22 V
60 22 V
60 21 V
60 22 V
61 21 V
60 22 V
60 21 V
60 22 V
60 21 V
61 22 V
60 22 V
60 21 V
0.800 UP
stroke
LT0
2167 2736 M
0 18 V
-31 -18 R
62 0 V
-62 18 R
62 0 V
379 56 R
0 15 V
-31 -15 R
62 0 V
-62 15 R
62 0 V
379 70 R
0 18 V
-31 -18 R
62 0 V
-62 18 R
62 0 V
379 71 R
0 16 V
-31 -16 R
62 0 V
-62 16 R
62 0 V
379 98 R
0 19 V
-31 -19 R
62 0 V
-62 19 R
62 0 V
379 65 R
0 22 V
-31 -22 R
62 0 V
-62 22 R
62 0 V
378 80 R
0 20 V
-31 -20 R
62 0 V
-62 20 R
62 0 V
2167 2745 Box
2577 2818 Box
2987 2904 Box
3397 2992 Box
3807 3107 Box
4217 3193 Box
4626 3294 Box
0.500 UL
LTb
900 2704 M
60 4 V
60 4 V
61 5 V
60 5 V
60 5 V
60 6 V
60 5 V
61 6 V
60 6 V
60 6 V
60 6 V
60 7 V
61 6 V
60 7 V
60 7 V
60 8 V
60 7 V
61 8 V
60 7 V
60 8 V
60 9 V
60 8 V
61 8 V
60 9 V
60 9 V
60 9 V
60 9 V
61 9 V
60 10 V
60 9 V
60 10 V
60 10 V
61 10 V
60 10 V
60 10 V
60 11 V
60 11 V
61 10 V
60 11 V
60 11 V
60 11 V
60 12 V
61 11 V
60 12 V
60 12 V
60 11 V
60 12 V
61 12 V
60 13 V
60 12 V
60 12 V
61 13 V
60 13 V
60 12 V
60 13 V
60 13 V
61 13 V
60 13 V
60 14 V
60 13 V
60 14 V
61 13 V
60 14 V
60 14 V
60 14 V
60 14 V
61 14 V
60 14 V
60 14 V
60 15 V
60 14 V
61 15 V
60 14 V
60 15 V
60 15 V
60 15 V
61 14 V
60 15 V
60 16 V
60 15 V
60 15 V
61 15 V
60 16 V
60 15 V
60 16 V
60 15 V
61 16 V
60 16 V
60 15 V
60 16 V
60 16 V
61 16 V
60 16 V
60 16 V
60 16 V
60 17 V
61 16 V
60 16 V
60 17 V
0.800 UP
stroke
LT0
2167 3530 M
0 30 V
-31 -30 R
62 0 V
-62 30 R
62 0 V
-31 101 R
0 30 V
-31 -30 R
62 0 V
-62 30 R
62 0 V
379 -13 R
0 32 V
-31 -32 R
62 0 V
-62 32 R
62 0 V
-31 -179 R
0 33 V
-31 -33 R
62 0 V
-62 33 R
62 0 V
379 27 R
0 38 V
-31 -38 R
62 0 V
-62 38 R
62 0 V
-31 65 R
0 37 V
-31 -37 R
62 0 V
-62 37 R
62 0 V
379 -34 R
0 94 V
-31 -94 R
62 0 V
-62 94 R
62 0 V
-31 -38 R
0 31 V
-31 -31 R
62 0 V
-62 31 R
62 0 V
379 -70 R
0 27 V
-31 -27 R
62 0 V
-62 27 R
62 0 V
-31 75 R
0 36 V
-31 -36 R
62 0 V
-62 36 R
62 0 V
379 -43 R
0 35 V
-31 -35 R
62 0 V
-62 35 R
62 0 V
-31 88 R
0 36 V
-31 -36 R
62 0 V
-62 36 R
62 0 V
378 -63 R
0 37 V
-31 -37 R
62 0 V
-62 37 R
62 0 V
-31 33 R
0 34 V
-31 -34 R
62 0 V
-62 34 R
62 0 V
2167 3545 Circle
2167 3676 Circle
2577 3694 Circle
2577 3548 Circle
2987 3610 Circle
2987 3713 Circle
3397 3744 Circle
3397 3768 Circle
3807 3727 Circle
3807 3834 Circle
4217 3826 Circle
4217 3950 Circle
4626 3923 Circle
4626 3992 Circle
0.800 UP
0.500 UL
LT2
2167 3399 M
0 85 V
-31 -85 R
62 0 V
-62 85 R
62 0 V
-31 241 R
0 126 V
-31 -126 R
62 0 V
-62 126 R
62 0 V
379 -149 R
0 33 V
-31 -33 R
62 0 V
-62 33 R
62 0 V
-31 146 R
0 46 V
-31 -46 R
62 0 V
-62 46 R
62 0 V
379 -226 R
0 28 V
-31 -28 R
62 0 V
-62 28 R
62 0 V
-31 166 R
0 33 V
-31 -33 R
62 0 V
-62 33 R
62 0 V
379 -207 R
0 29 V
-31 -29 R
62 0 V
-62 29 R
62 0 V
-31 80 R
0 32 V
-31 -32 R
62 0 V
-62 32 R
62 0 V
379 -39 R
0 35 V
-31 -35 R
62 0 V
-62 35 R
62 0 V
-31 39 R
0 38 V
-31 -38 R
62 0 V
-62 38 R
62 0 V
379 -31 R
0 44 V
-31 -44 R
62 0 V
-62 44 R
62 0 V
-31 13 R
0 37 V
-31 -37 R
62 0 V
-62 37 R
62 0 V
378 -6 R
0 37 V
-31 -37 R
62 0 V
-62 37 R
62 0 V
-31 7 R
0 40 V
-31 -40 R
62 0 V
-62 40 R
62 0 V
2167 3441 TriU
2167 3788 TriU
2577 3718 TriU
2577 3904 TriU
2987 3715 TriU
2987 3911 TriU
3397 3735 TriU
3397 3846 TriU
3807 3840 TriU
3807 3916 TriU
4217 3926 TriU
4217 3979 TriU
4626 4010 TriU
4626 4056 TriU
0.500 UL
LTb
900 3576 M
60 3 V
60 4 V
61 3 V
60 3 V
60 4 V
60 4 V
60 4 V
61 4 V
60 4 V
60 5 V
60 4 V
60 5 V
61 5 V
60 5 V
60 5 V
60 5 V
60 5 V
61 6 V
60 5 V
60 6 V
60 6 V
60 6 V
61 6 V
60 7 V
60 6 V
60 7 V
60 6 V
61 7 V
60 7 V
60 7 V
60 7 V
60 8 V
61 7 V
60 8 V
60 8 V
60 7 V
60 8 V
61 8 V
60 9 V
60 8 V
60 8 V
60 9 V
61 9 V
60 8 V
60 9 V
60 9 V
60 10 V
61 9 V
60 9 V
60 10 V
60 9 V
61 10 V
60 10 V
60 10 V
60 10 V
60 10 V
61 10 V
60 10 V
60 11 V
60 10 V
60 11 V
61 10 V
60 11 V
60 11 V
60 11 V
60 11 V
61 12 V
60 11 V
60 11 V
60 12 V
60 11 V
61 12 V
60 12 V
60 11 V
60 12 V
60 12 V
61 12 V
60 13 V
60 12 V
60 12 V
60 13 V
61 12 V
60 13 V
60 12 V
60 13 V
60 13 V
61 13 V
60 13 V
60 13 V
60 13 V
60 13 V
61 13 V
60 14 V
60 13 V
60 14 V
60 13 V
61 14 V
60 13 V
60 14 V
stroke
1.500 UL
LT1
3669 3518 M
60 -2 V
61 -1 V
60 0 V
60 0 V
60 1 V
61 2 V
60 2 V
60 2 V
60 4 V
60 3 V
61 4 V
60 5 V
60 4 V
60 6 V
60 5 V
61 6 V
60 7 V
60 7 V
60 7 V
60 7 V
61 8 V
60 8 V
60 8 V
60 9 V
60 9 V
61 9 V
60 10 V
60 9 V
60 10 V
60 11 V
61 10 V
60 11 V
60 10 V
60 11 V
60 12 V
61 11 V
60 12 V
60 12 V
60 12 V
60 12 V
61 12 V
60 13 V
60 12 V
60 13 V
60 13 V
61 13 V
60 14 V
60 13 V
60 14 V
60 13 V
61 14 V
60 14 V
60 14 V
3669 2688 M
60 35 V
61 34 V
60 32 V
60 31 V
60 30 V
61 29 V
60 28 V
60 27 V
60 27 V
60 25 V
61 25 V
60 25 V
60 23 V
60 23 V
60 23 V
61 23 V
60 21 V
60 22 V
60 21 V
60 21 V
61 20 V
60 21 V
60 20 V
60 20 V
60 19 V
61 20 V
60 19 V
60 19 V
60 19 V
60 19 V
61 18 V
60 19 V
60 18 V
60 19 V
60 18 V
61 18 V
60 18 V
60 18 V
60 18 V
60 18 V
61 18 V
60 18 V
60 18 V
60 18 V
60 17 V
61 18 V
60 18 V
60 18 V
60 17 V
60 18 V
6740 3787 L
60 17 V
60 18 V
stroke
1.000 UL
LTb
900 4800 M
900 600 L
5960 0 V
0 4200 V
-5960 0 V
1.000 UP
stroke
grestore
end
showpage
  }}%
  \put(3880,100){\makebox(0,0){\strut{}$\sqrt{\sigma} l$}}%
  \put(200,2700){%
  \special{ps: gsave currentpoint currentpoint translate
 0 rotate neg exch neg exch translate}%
  \makebox(0,0){\strut{}$E / \sqrt{\sigma}$}%
  \special{ps: currentpoint grestore moveto}%
  }%
  \put(6860,400){\makebox(0,0){\strut{} 6}}%
  \put(5668,400){\makebox(0,0){\strut{} 5}}%
  \put(4476,400){\makebox(0,0){\strut{} 4}}%
  \put(3284,400){\makebox(0,0){\strut{} 3}}%
  \put(2092,400){\makebox(0,0){\strut{} 2}}%
  \put(900,400){\makebox(0,0){\strut{} 1}}%
  \put(780,4800){\makebox(0,0)[r]{\strut{} 10}}%
  \put(780,3960){\makebox(0,0)[r]{\strut{} 8}}%
  \put(780,3120){\makebox(0,0)[r]{\strut{} 6}}%
  \put(780,2280){\makebox(0,0)[r]{\strut{} 4}}%
  \put(780,1440){\makebox(0,0)[r]{\strut{} 2}}%
  \put(780,600){\makebox(0,0)[r]{\strut{} 0}}%
\end{picture}%
\endgroup
 

%% file: mom4.tex
\begingroup%
  \makeatletter%
  \newcommand{\GNUPLOTspecial}{%
    \@sanitize\catcode`\%=14\relax\special}%
  \setlength{\unitlength}{0.1bp}%
\begin{picture}(3600,2160)(0,0)%
{\GNUPLOTspecial{"
/gnudict 256 dict def
gnudict begin
/Color true def
/Solid true def
/gnulinewidth 5.000 def
/userlinewidth gnulinewidth def
/vshift -33 def
/dl {10.0 mul} def
/hpt_ 31.5 def
/vpt_ 31.5 def
/hpt hpt_ def
/vpt vpt_ def
/Rounded false def
/M {moveto} bind def
/L {lineto} bind def
/R {rmoveto} bind def
/V {rlineto} bind def
/N {newpath moveto} bind def
/C {setrgbcolor} bind def
/f {rlineto fill} bind def
/vpt2 vpt 2 mul def
/hpt2 hpt 2 mul def
/Lshow { currentpoint stroke M
  0 vshift R show } def
/Rshow { currentpoint stroke M
  dup stringwidth pop neg vshift R show } def
/Cshow { currentpoint stroke M
  dup stringwidth pop -2 div vshift R show } def
/UP { dup vpt_ mul /vpt exch def hpt_ mul /hpt exch def
  /hpt2 hpt 2 mul def /vpt2 vpt 2 mul def } def
/DL { Color {setrgbcolor Solid {pop []} if 0 setdash }
 {pop pop pop 0 setgray Solid {pop []} if 0 setdash} ifelse } def
/BL { stroke userlinewidth 2 mul setlinewidth
      Rounded { 1 setlinejoin 1 setlinecap } if } def
/AL { stroke userlinewidth 2 div setlinewidth
      Rounded { 1 setlinejoin 1 setlinecap } if } def
/UL { dup gnulinewidth mul /userlinewidth exch def
      dup 1 lt {pop 1} if 10 mul /udl exch def } def
/PL { stroke userlinewidth setlinewidth
      Rounded { 1 setlinejoin 1 setlinecap } if } def
/LTw { PL [] 1 setgray } def
/LTb { BL [] 0 0 0 DL } def
/LTa { AL [1 udl mul 2 udl mul] 0 setdash 0 0 0 setrgbcolor } def
/LT0 { PL [] 1 0 0 DL } def
/LT1 { PL [4 dl 2 dl] 0 1 0 DL } def
/LT2 { PL [2 dl 3 dl] 0 0 1 DL } def
/LT3 { PL [1 dl 1.5 dl] 1 0 1 DL } def
/LT4 { PL [5 dl 2 dl 1 dl 2 dl] 0 1 1 DL } def
/LT5 { PL [4 dl 3 dl 1 dl 3 dl] 1 1 0 DL } def
/LT6 { PL [2 dl 2 dl 2 dl 4 dl] 0 0 0 DL } def
/LT7 { PL [2 dl 2 dl 2 dl 2 dl 2 dl 4 dl] 1 0.3 0 DL } def
/LT8 { PL [2 dl 2 dl 2 dl 2 dl 2 dl 2 dl 2 dl 4 dl] 0.5 0.5 0.5 DL } def
/Pnt { stroke [] 0 setdash
   gsave 1 setlinecap M 0 0 V stroke grestore } def
/Dia { stroke [] 0 setdash 2 copy vpt add M
  hpt neg vpt neg V hpt vpt neg V
  hpt vpt V hpt neg vpt V closepath stroke
  Pnt } def
/Pls { stroke [] 0 setdash vpt sub M 0 vpt2 V
  currentpoint stroke M
  hpt neg vpt neg R hpt2 0 V stroke
  } def
/Box { stroke [] 0 setdash 2 copy exch hpt sub exch vpt add M
  0 vpt2 neg V hpt2 0 V 0 vpt2 V
  hpt2 neg 0 V closepath stroke
  Pnt } def
/Crs { stroke [] 0 setdash exch hpt sub exch vpt add M
  hpt2 vpt2 neg V currentpoint stroke M
  hpt2 neg 0 R hpt2 vpt2 V stroke } def
/TriU { stroke [] 0 setdash 2 copy vpt 1.12 mul add M
  hpt neg vpt -1.62 mul V
  hpt 2 mul 0 V
  hpt neg vpt 1.62 mul V closepath stroke
  Pnt  } def
/Star { 2 copy Pls Crs } def
/BoxF { stroke [] 0 setdash exch hpt sub exch vpt add M
  0 vpt2 neg V  hpt2 0 V  0 vpt2 V
  hpt2 neg 0 V  closepath fill } def
/TriUF { stroke [] 0 setdash vpt 1.12 mul add M
  hpt neg vpt -1.62 mul V
  hpt 2 mul 0 V
  hpt neg vpt 1.62 mul V closepath fill } def
/TriD { stroke [] 0 setdash 2 copy vpt 1.12 mul sub M
  hpt neg vpt 1.62 mul V
  hpt 2 mul 0 V
  hpt neg vpt -1.62 mul V closepath stroke
  Pnt  } def
/TriDF { stroke [] 0 setdash vpt 1.12 mul sub M
  hpt neg vpt 1.62 mul V
  hpt 2 mul 0 V
  hpt neg vpt -1.62 mul V closepath fill} def
/DiaF { stroke [] 0 setdash vpt add M
  hpt neg vpt neg V hpt vpt neg V
  hpt vpt V hpt neg vpt V closepath fill } def
/Pent { stroke [] 0 setdash 2 copy gsave
  translate 0 hpt M 4 {72 rotate 0 hpt L} repeat
  closepath stroke grestore Pnt } def
/PentF { stroke [] 0 setdash gsave
  translate 0 hpt M 4 {72 rotate 0 hpt L} repeat
  closepath fill grestore } def
/Circle { stroke [] 0 setdash 2 copy
  hpt 0 360 arc stroke Pnt } def
/CircleF { stroke [] 0 setdash hpt 0 360 arc fill } def
/C0 { BL [] 0 setdash 2 copy moveto vpt 90 450  arc } bind def
/C1 { BL [] 0 setdash 2 copy        moveto
       2 copy  vpt 0 90 arc closepath fill
               vpt 0 360 arc closepath } bind def
/C2 { BL [] 0 setdash 2 copy moveto
       2 copy  vpt 90 180 arc closepath fill
               vpt 0 360 arc closepath } bind def
/C3 { BL [] 0 setdash 2 copy moveto
       2 copy  vpt 0 180 arc closepath fill
               vpt 0 360 arc closepath } bind def
/C4 { BL [] 0 setdash 2 copy moveto
       2 copy  vpt 180 270 arc closepath fill
               vpt 0 360 arc closepath } bind def
/C5 { BL [] 0 setdash 2 copy moveto
       2 copy  vpt 0 90 arc
       2 copy moveto
       2 copy  vpt 180 270 arc closepath fill
               vpt 0 360 arc } bind def
/C6 { BL [] 0 setdash 2 copy moveto
      2 copy  vpt 90 270 arc closepath fill
              vpt 0 360 arc closepath } bind def
/C7 { BL [] 0 setdash 2 copy moveto
      2 copy  vpt 0 270 arc closepath fill
              vpt 0 360 arc closepath } bind def
/C8 { BL [] 0 setdash 2 copy moveto
      2 copy vpt 270 360 arc closepath fill
              vpt 0 360 arc closepath } bind def
/C9 { BL [] 0 setdash 2 copy moveto
      2 copy  vpt 270 450 arc closepath fill
              vpt 0 360 arc closepath } bind def
/C10 { BL [] 0 setdash 2 copy 2 copy moveto vpt 270 360 arc closepath fill
       2 copy moveto
       2 copy vpt 90 180 arc closepath fill
               vpt 0 360 arc closepath } bind def
/C11 { BL [] 0 setdash 2 copy moveto
       2 copy  vpt 0 180 arc closepath fill
       2 copy moveto
       2 copy  vpt 270 360 arc closepath fill
               vpt 0 360 arc closepath } bind def
/C12 { BL [] 0 setdash 2 copy moveto
       2 copy  vpt 180 360 arc closepath fill
               vpt 0 360 arc closepath } bind def
/C13 { BL [] 0 setdash  2 copy moveto
       2 copy  vpt 0 90 arc closepath fill
       2 copy moveto
       2 copy  vpt 180 360 arc closepath fill
               vpt 0 360 arc closepath } bind def
/C14 { BL [] 0 setdash 2 copy moveto
       2 copy  vpt 90 360 arc closepath fill
               vpt 0 360 arc } bind def
/C15 { BL [] 0 setdash 2 copy vpt 0 360 arc closepath fill
               vpt 0 360 arc closepath } bind def
/Rec   { newpath 4 2 roll moveto 1 index 0 rlineto 0 exch rlineto
       neg 0 rlineto closepath } bind def
/Square { dup Rec } bind def
/Bsquare { vpt sub exch vpt sub exch vpt2 Square } bind def
/S0 { BL [] 0 setdash 2 copy moveto 0 vpt rlineto BL Bsquare } bind def
/S1 { BL [] 0 setdash 2 copy vpt Square fill Bsquare } bind def
/S2 { BL [] 0 setdash 2 copy exch vpt sub exch vpt Square fill Bsquare } bind def
/S3 { BL [] 0 setdash 2 copy exch vpt sub exch vpt2 vpt Rec fill Bsquare } bind def
/S4 { BL [] 0 setdash 2 copy exch vpt sub exch vpt sub vpt Square fill Bsquare } bind def
/S5 { BL [] 0 setdash 2 copy 2 copy vpt Square fill
       exch vpt sub exch vpt sub vpt Square fill Bsquare } bind def
/S6 { BL [] 0 setdash 2 copy exch vpt sub exch vpt sub vpt vpt2 Rec fill Bsquare } bind def
/S7 { BL [] 0 setdash 2 copy exch vpt sub exch vpt sub vpt vpt2 Rec fill
       2 copy vpt Square fill
       Bsquare } bind def
/S8 { BL [] 0 setdash 2 copy vpt sub vpt Square fill Bsquare } bind def
/S9 { BL [] 0 setdash 2 copy vpt sub vpt vpt2 Rec fill Bsquare } bind def
/S10 { BL [] 0 setdash 2 copy vpt sub vpt Square fill 2 copy exch vpt sub exch vpt Square fill
       Bsquare } bind def
/S11 { BL [] 0 setdash 2 copy vpt sub vpt Square fill 2 copy exch vpt sub exch vpt2 vpt Rec fill
       Bsquare } bind def
/S12 { BL [] 0 setdash 2 copy exch vpt sub exch vpt sub vpt2 vpt Rec fill Bsquare } bind def
/S13 { BL [] 0 setdash 2 copy exch vpt sub exch vpt sub vpt2 vpt Rec fill
       2 copy vpt Square fill Bsquare } bind def
/S14 { BL [] 0 setdash 2 copy exch vpt sub exch vpt sub vpt2 vpt Rec fill
       2 copy exch vpt sub exch vpt Square fill Bsquare } bind def
/S15 { BL [] 0 setdash 2 copy Bsquare fill Bsquare } bind def
/D0 { gsave translate 45 rotate 0 0 S0 stroke grestore } bind def
/D1 { gsave translate 45 rotate 0 0 S1 stroke grestore } bind def
/D2 { gsave translate 45 rotate 0 0 S2 stroke grestore } bind def
/D3 { gsave translate 45 rotate 0 0 S3 stroke grestore } bind def
/D4 { gsave translate 45 rotate 0 0 S4 stroke grestore } bind def
/D5 { gsave translate 45 rotate 0 0 S5 stroke grestore } bind def
/D6 { gsave translate 45 rotate 0 0 S6 stroke grestore } bind def
/D7 { gsave translate 45 rotate 0 0 S7 stroke grestore } bind def
/D8 { gsave translate 45 rotate 0 0 S8 stroke grestore } bind def
/D9 { gsave translate 45 rotate 0 0 S9 stroke grestore } bind def
/D10 { gsave translate 45 rotate 0 0 S10 stroke grestore } bind def
/D11 { gsave translate 45 rotate 0 0 S11 stroke grestore } bind def
/D12 { gsave translate 45 rotate 0 0 S12 stroke grestore } bind def
/D13 { gsave translate 45 rotate 0 0 S13 stroke grestore } bind def
/D14 { gsave translate 45 rotate 0 0 S14 stroke grestore } bind def
/D15 { gsave translate 45 rotate 0 0 S15 stroke grestore } bind def
/DiaE { stroke [] 0 setdash vpt add M
  hpt neg vpt neg V hpt vpt neg V
  hpt vpt V hpt neg vpt V closepath stroke } def
/BoxE { stroke [] 0 setdash exch hpt sub exch vpt add M
  0 vpt2 neg V hpt2 0 V 0 vpt2 V
  hpt2 neg 0 V closepath stroke } def
/TriUE { stroke [] 0 setdash vpt 1.12 mul add M
  hpt neg vpt -1.62 mul V
  hpt 2 mul 0 V
  hpt neg vpt 1.62 mul V closepath stroke } def
/TriDE { stroke [] 0 setdash vpt 1.12 mul sub M
  hpt neg vpt 1.62 mul V
  hpt 2 mul 0 V
  hpt neg vpt -1.62 mul V closepath stroke } def
/PentE { stroke [] 0 setdash gsave
  translate 0 hpt M 4 {72 rotate 0 hpt L} repeat
  closepath stroke grestore } def
/CircE { stroke [] 0 setdash 
  hpt 0 360 arc stroke } def
/Opaque { gsave closepath 1 setgray fill grestore 0 setgray closepath } def
/DiaW { stroke [] 0 setdash vpt add M
  hpt neg vpt neg V hpt vpt neg V
  hpt vpt V hpt neg vpt V Opaque stroke } def
/BoxW { stroke [] 0 setdash exch hpt sub exch vpt add M
  0 vpt2 neg V hpt2 0 V 0 vpt2 V
  hpt2 neg 0 V Opaque stroke } def
/TriUW { stroke [] 0 setdash vpt 1.12 mul add M
  hpt neg vpt -1.62 mul V
  hpt 2 mul 0 V
  hpt neg vpt 1.62 mul V Opaque stroke } def
/TriDW { stroke [] 0 setdash vpt 1.12 mul sub M
  hpt neg vpt 1.62 mul V
  hpt 2 mul 0 V
  hpt neg vpt -1.62 mul V Opaque stroke } def
/PentW { stroke [] 0 setdash gsave
  translate 0 hpt M 4 {72 rotate 0 hpt L} repeat
  Opaque stroke grestore } def
/CircW { stroke [] 0 setdash 
  hpt 0 360 arc Opaque stroke } def
/BoxFill { gsave Rec 1 setgray fill grestore } def
/BoxColFill {
  gsave Rec
  /Fillden exch def
  currentrgbcolor
  /ColB exch def /ColG exch def /ColR exch def
  /ColR ColR Fillden mul Fillden sub 1 add def
  /ColG ColG Fillden mul Fillden sub 1 add def
  /ColB ColB Fillden mul Fillden sub 1 add def
  ColR ColG ColB setrgbcolor
  fill grestore } def
%
%
/PatternFill { gsave /PFa [ 9 2 roll ] def
    PFa 0 get PFa 2 get 2 div add PFa 1 get PFa 3 get 2 div add translate
    PFa 2 get -2 div PFa 3 get -2 div PFa 2 get PFa 3 get Rec
    gsave 1 setgray fill grestore clip
    currentlinewidth 0.5 mul setlinewidth
    /PFs PFa 2 get dup mul PFa 3 get dup mul add sqrt def
    0 0 M PFa 5 get rotate PFs -2 div dup translate
	0 1 PFs PFa 4 get div 1 add floor cvi
	{ PFa 4 get mul 0 M 0 PFs V } for
    0 PFa 6 get ne {
	0 1 PFs PFa 4 get div 1 add floor cvi
	{ PFa 4 get mul 0 2 1 roll M PFs 0 V } for
    } if
    stroke grestore } def
/Symbol-Oblique /Symbol findfont [1 0 .167 1 0 0] makefont
dup length dict begin {1 index /FID eq {pop pop} {def} ifelse} forall
currentdict end definefont pop
end
gnudict begin
gsave
0 0 translate
0.100 0.100 scale
0 setgray
newpath
1.000 UL
LTb
350 300 M
63 0 V
3037 0 R
-63 0 V
1.000 UL
LTb
350 496 M
63 0 V
3037 0 R
-63 0 V
1.000 UL
LTb
350 691 M
63 0 V
3037 0 R
-63 0 V
1.000 UL
LTb
350 887 M
63 0 V
3037 0 R
-63 0 V
1.000 UL
LTb
350 1082 M
63 0 V
3037 0 R
-63 0 V
1.000 UL
LTb
350 1278 M
63 0 V
3037 0 R
-63 0 V
1.000 UL
LTb
350 1473 M
63 0 V
3037 0 R
-63 0 V
1.000 UL
LTb
350 1669 M
63 0 V
3037 0 R
-63 0 V
1.000 UL
LTb
350 1864 M
63 0 V
3037 0 R
-63 0 V
1.000 UL
LTb
350 2060 M
63 0 V
3037 0 R
-63 0 V
1.000 UL
LTb
350 300 M
0 63 V
0 1697 R
0 -63 V
1.000 UL
LTb
970 300 M
0 63 V
0 1697 R
0 -63 V
1.000 UL
LTb
1590 300 M
0 63 V
0 1697 R
0 -63 V
1.000 UL
LTb
2210 300 M
0 63 V
0 1697 R
0 -63 V
1.000 UL
LTb
2830 300 M
0 63 V
0 1697 R
0 -63 V
1.000 UL
LTb
3450 300 M
0 63 V
0 1697 R
0 -63 V
1.000 UL
LTb
1.000 UL
LTb
350 300 M
3100 0 V
0 1760 V
-3100 0 V
350 300 L
LTb
LTb
1.000 UP
0.400 UP
0.500 UL
LT2
LTb
LT2
3087 1203 M
263 0 V
-263 31 R
0 -62 V
263 62 R
0 -62 V
458 1067 M
0 8 V
-31 -8 R
62 0 V
-62 8 R
62 0 V
1263 128 R
0 7 V
-31 -7 R
62 0 V
-62 7 R
62 0 V
1264 140 R
0 7 V
-31 -7 R
62 0 V
-62 7 R
62 0 V
458 1071 Pls
1752 1206 Pls
3047 1353 Pls
3218 1203 Pls
0.400 UP
0.500 UL
LT0
LTb
LT0
3087 1123 M
263 0 V
-263 31 R
0 -62 V
263 62 R
0 -62 V
458 1542 M
0 11 V
-31 -11 R
62 0 V
-62 11 R
62 0 V
1263 57 R
0 14 V
-31 -14 R
62 0 V
-62 14 R
62 0 V
1264 103 R
0 16 V
-31 -16 R
62 0 V
-62 16 R
62 0 V
458 1547 Crs
1752 1617 Crs
3047 1735 Crs
3218 1123 Crs
0.400 UP
0.500 UL
LT2
LTb
LT2
3087 1043 M
263 0 V
-263 31 R
0 -62 V
263 62 R
0 -62 V
458 1556 M
0 13 V
-31 -13 R
62 0 V
-62 13 R
62 0 V
1263 55 R
0 12 V
-31 -12 R
62 0 V
-62 12 R
62 0 V
1264 89 R
0 17 V
-31 -17 R
62 0 V
-62 17 R
62 0 V
458 1563 Star
1752 1630 Star
3047 1734 Star
3218 1043 Star
0.400 UP
0.500 UL
LT0
LTb
LT0
3087 963 M
263 0 V
-263 31 R
0 -62 V
263 62 R
0 -62 V
458 1310 M
0 31 V
-31 -31 R
62 0 V
-62 31 R
62 0 V
1263 84 R
0 17 V
-31 -17 R
62 0 V
-62 17 R
62 0 V
1264 118 R
0 15 V
-31 -15 R
62 0 V
-62 15 R
62 0 V
458 1326 Box
1752 1433 Box
3047 1568 Box
3218 963 Box
0.400 UP
0.500 UL
LT2
LTb
LT2
3087 883 M
263 0 V
-263 31 R
0 -62 V
263 62 R
0 -62 V
458 1226 M
0 27 V
-31 -27 R
62 0 V
-62 27 R
62 0 V
1263 160 R
0 15 V
-31 -15 R
62 0 V
-62 15 R
62 0 V
1264 123 R
0 13 V
-31 -13 R
62 0 V
-62 13 R
62 0 V
458 1240 BoxF
1752 1420 BoxF
3047 1557 BoxF
3218 883 BoxF
0.400 UP
0.500 UL
LT0
LTb
LT0
3087 803 M
263 0 V
-263 31 R
0 -62 V
263 62 R
0 -62 V
458 1737 M
0 38 V
-31 -38 R
62 0 V
-62 38 R
62 0 V
-31 -62 R
0 45 V
-31 -45 R
62 0 V
-62 45 R
62 0 V
1263 46 R
0 25 V
-31 -25 R
62 0 V
-62 25 R
62 0 V
-31 -23 R
0 34 V
-31 -34 R
62 0 V
-62 34 R
62 0 V
1264 75 R
0 23 V
-31 -23 R
62 0 V
-62 23 R
62 0 V
-31 0 R
0 20 V
-31 -20 R
62 0 V
-62 20 R
62 0 V
458 1756 Circle
458 1736 Circle
1752 1817 Circle
1752 1823 Circle
3047 1926 Circle
3047 1948 Circle
3218 803 Circle
0.400 UP
0.500 UL
LT2
LTb
LT2
3087 723 M
263 0 V
-263 31 R
0 -62 V
263 62 R
0 -62 V
458 1711 M
0 44 V
-31 -44 R
62 0 V
-62 44 R
62 0 V
1263 67 R
0 31 V
-31 -31 R
62 0 V
-62 31 R
62 0 V
1264 64 R
0 26 V
-31 -26 R
62 0 V
-62 26 R
62 0 V
458 1733 CircleF
1752 1838 CircleF
3047 1930 CircleF
3218 723 CircleF
0.500 UL
LTb
LTb
LTb
3087 643 M
263 0 V
350 1070 M
31 3 V
32 2 V
31 3 V
31 3 V
32 2 V
31 3 V
31 2 V
32 3 V
31 3 V
31 3 V
31 2 V
32 3 V
31 3 V
31 3 V
32 3 V
31 2 V
31 3 V
32 3 V
31 3 V
31 3 V
32 3 V
31 3 V
31 3 V
32 3 V
31 3 V
31 3 V
31 3 V
32 3 V
31 3 V
31 3 V
32 3 V
31 3 V
31 4 V
32 3 V
31 3 V
31 3 V
32 3 V
31 4 V
31 3 V
32 3 V
31 3 V
31 4 V
31 3 V
32 3 V
31 4 V
31 3 V
32 3 V
31 4 V
31 3 V
32 4 V
31 3 V
31 4 V
32 3 V
31 3 V
31 4 V
32 3 V
31 4 V
31 3 V
31 4 V
32 4 V
31 3 V
31 4 V
32 3 V
31 4 V
31 3 V
32 4 V
31 4 V
31 3 V
32 4 V
31 4 V
31 3 V
32 4 V
31 4 V
31 3 V
31 4 V
32 4 V
31 4 V
31 3 V
32 4 V
31 4 V
31 4 V
32 3 V
31 4 V
31 4 V
32 4 V
31 4 V
31 4 V
32 3 V
31 4 V
31 4 V
31 4 V
32 4 V
31 4 V
31 4 V
32 4 V
31 3 V
31 4 V
32 4 V
31 4 V
0.500 UL
LT1
LTb
LT1
3087 563 M
263 0 V
350 1547 M
31 1 V
32 2 V
31 2 V
31 1 V
32 2 V
31 1 V
31 2 V
32 2 V
31 1 V
31 2 V
31 2 V
32 2 V
31 1 V
31 2 V
32 2 V
31 2 V
31 2 V
32 2 V
31 1 V
31 2 V
32 2 V
31 2 V
31 2 V
32 2 V
31 2 V
31 2 V
31 2 V
32 2 V
31 2 V
31 2 V
32 2 V
31 2 V
31 2 V
32 2 V
31 2 V
31 2 V
32 3 V
31 2 V
31 2 V
32 2 V
31 2 V
31 3 V
31 2 V
32 2 V
31 2 V
31 3 V
32 2 V
31 2 V
31 3 V
32 2 V
31 2 V
31 3 V
32 2 V
31 2 V
31 3 V
32 2 V
31 3 V
31 2 V
31 3 V
32 2 V
31 2 V
31 3 V
32 3 V
31 2 V
31 3 V
32 2 V
31 3 V
31 2 V
32 3 V
31 2 V
31 3 V
32 3 V
31 2 V
31 3 V
31 3 V
32 2 V
31 3 V
31 3 V
32 2 V
31 3 V
31 3 V
32 3 V
31 2 V
31 3 V
32 3 V
31 3 V
31 3 V
32 2 V
31 3 V
31 3 V
31 3 V
32 3 V
31 3 V
31 3 V
32 2 V
31 3 V
31 3 V
32 3 V
31 3 V
0.500 UL
LT3
LTb
LT3
3087 483 M
263 0 V
350 1336 M
31 2 V
32 2 V
31 2 V
31 2 V
32 2 V
31 2 V
31 2 V
32 2 V
31 2 V
31 2 V
31 2 V
32 2 V
31 2 V
31 2 V
32 3 V
31 2 V
31 2 V
32 2 V
31 2 V
31 3 V
32 2 V
31 2 V
31 2 V
32 3 V
31 2 V
31 2 V
31 3 V
32 2 V
31 2 V
31 3 V
32 2 V
31 3 V
31 2 V
32 3 V
31 2 V
31 3 V
32 2 V
31 3 V
31 2 V
32 3 V
31 2 V
31 3 V
31 3 V
32 2 V
31 3 V
31 3 V
32 2 V
31 3 V
31 3 V
32 2 V
31 3 V
31 3 V
32 3 V
31 3 V
31 2 V
32 3 V
31 3 V
31 3 V
31 3 V
32 3 V
31 2 V
31 3 V
32 3 V
31 3 V
31 3 V
32 3 V
31 3 V
31 3 V
32 3 V
31 3 V
31 3 V
32 3 V
31 3 V
31 3 V
31 3 V
32 3 V
31 3 V
31 3 V
32 3 V
31 4 V
31 3 V
32 3 V
31 3 V
31 3 V
32 3 V
31 4 V
31 3 V
32 3 V
31 3 V
31 3 V
31 4 V
32 3 V
31 3 V
31 4 V
32 3 V
31 3 V
31 3 V
32 4 V
31 3 V
0.500 UL
LT4
LTb
LT4
3087 403 M
263 0 V
350 1727 M
31 1 V
32 1 V
31 2 V
31 1 V
32 2 V
31 1 V
31 1 V
32 2 V
31 1 V
31 2 V
31 1 V
32 2 V
31 2 V
31 1 V
32 2 V
31 1 V
31 2 V
32 1 V
31 2 V
31 2 V
32 1 V
31 2 V
31 2 V
32 2 V
31 1 V
31 2 V
31 2 V
32 2 V
31 1 V
31 2 V
32 2 V
31 2 V
31 2 V
32 1 V
31 2 V
31 2 V
32 2 V
31 2 V
31 2 V
32 2 V
31 2 V
31 2 V
31 2 V
32 2 V
31 2 V
31 2 V
32 2 V
31 2 V
31 2 V
32 2 V
31 2 V
31 2 V
32 2 V
31 2 V
31 3 V
32 2 V
31 2 V
31 2 V
31 2 V
32 3 V
31 2 V
31 2 V
32 2 V
31 2 V
31 3 V
32 2 V
31 2 V
31 3 V
32 2 V
31 2 V
31 3 V
32 2 V
31 2 V
31 3 V
31 2 V
32 2 V
31 3 V
31 2 V
32 3 V
31 2 V
31 3 V
32 2 V
31 3 V
31 2 V
32 3 V
31 2 V
31 3 V
32 2 V
31 3 V
31 2 V
31 3 V
32 3 V
31 2 V
31 3 V
32 2 V
31 3 V
31 3 V
32 2 V
31 3 V
1.000 UL
LTb
350 300 M
3100 0 V
0 1760 V
-3100 0 V
350 300 L
1.000 UP
stroke
grestore
end
showpage
}}%
\put(3037,403){\makebox(0,0)[r]{\scriptsize$N_R=3, N_L=1, q=2, w=1$}}%
\put(3037,483){\makebox(0,0)[r]{\scriptsize$N_R=2, N_L=0, q=2, w=1$}}%
\put(3037,563){\makebox(0,0)[r]{\scriptsize$N_R=2, N_L=1, q=1, w=1$}}%
\put(3037,643){\makebox(0,0)[r]{\scriptsize$N_R=1, N_L=0, q=1, w=1$}}%
\put(3037,723){\makebox(0,0)[r]{\scriptsize$1^{st}$ e.s for $q=2, P=-$}}%
\put(3037,803){\makebox(0,0)[r]{\scriptsize$1^{st}$ e.s for $q=2, P=+$}}%
\put(3037,883){\makebox(0,0)[r]{\scriptsize g.s for $q=2, P=-$}}%
\put(3037,963){\makebox(0,0)[r]{\scriptsize g.s for $q=2, P=+$}}%
\put(3037,1043){\makebox(0,0)[r]{\scriptsize$1^{st}$ e.s for $q=1, P=-$}}%
\put(3037,1123){\makebox(0,0)[r]{\scriptsize g.s for $q=1, P=+$}}%
\put(3037,1203){\makebox(0,0)[r]{\scriptsize g.s for $q=1, P=-$}}%
\put(1900,50){\makebox(0,0){$l \sqrt{\sigma}$}}%
\put(100,1180){%
\special{ps: gsave currentpoint currentpoint translate
270 rotate neg exch neg exch translate}%
\makebox(0,0)[b]{\shortstack{$\sqrt{ E^2/ \sigma - (2 \pi q/ \sqrt{\sigma} l)^2}$}}%
\special{ps: currentpoint grestore moveto}%
}%
\put(3450,200){\makebox(0,0){ 4.5}}%
\put(2830,200){\makebox(0,0){ 4}}%
\put(2210,200){\makebox(0,0){ 3.5}}%
\put(1590,200){\makebox(0,0){ 3}}%
\put(970,200){\makebox(0,0){ 2.5}}%
\put(350,200){\makebox(0,0){ 2}}%
\put(300,2060){\makebox(0,0)[r]{ 9}}%
\put(300,1864){\makebox(0,0)[r]{ 8}}%
\put(300,1669){\makebox(0,0)[r]{ 7}}%
\put(300,1473){\makebox(0,0)[r]{ 6}}%
\put(300,1278){\makebox(0,0)[r]{ 5}}%
\put(300,1082){\makebox(0,0)[r]{ 4}}%
\put(300,887){\makebox(0,0)[r]{ 3}}%
\put(300,691){\makebox(0,0)[r]{ 2}}%
\put(300,496){\makebox(0,0)[r]{ 1}}%
\put(300,300){\makebox(0,0)[r]{ 0}}%
\end{picture}%
\endgroup
 